\documentclass[prd,nofootinbib,superscriptaddress]{revtex4-2}
\usepackage{amsmath,amsfonts,amssymb}
\usepackage{graphicx}
\usepackage{dcolumn}
\usepackage[mathlines]{lineno}
\usepackage{float}
\usepackage{hyperref}

\usepackage[usenames,dvipsnames]{color}
\usepackage{etex}
\usepackage{booktabs}
\usepackage{adjustbox}
\usepackage{cleveref}

\begin{document}

\title{Fermion Masses and Mixing in Pati-Salam Unification with $S_3$ Modular Symmetry} 

\author{Mohamed Belfkir}
\email{m\_belfkir@uaeu.ac.ae}
\affiliation{Department of physics, United Arab Emirates University, Al-Ain, UAE}
\author{Mohamed Amin Loualidi}
\email{ma.loualidi@uaeu.ac.ae}
\affiliation{Department of physics, United Arab Emirates University, Al-Ain, UAE}
\author{Salah Nasri}
\email{snasri@uaeu.ac.ae, salah.nasri@cern.ch}
\affiliation{Department of physics, United Arab Emirates University, Al-Ain, UAE}

\begin{abstract}
Modular invariance has recently paved the way for promising new directions in flavor model building. Motivated by this development, we present in this work the first implementation of the $S_3$ modular symmetry within the Pati-Salam unification framework, addressing the flavor structure of quarks and leptons. Assigning left- and right-handed matter fields as $S_3$ doublets or singlets, we propose three benchmark models that achieve compelling fits to 16 observables including charged fermion mass ratios and flavor mixing parameters. Light neutrino masses arise via the type-I seesaw mechanism, and neutrino oscillation parameters are explored in light of the latest NuFIT v6.0 results. All models favor a normal neutrino mass ordering, with the atmospheric mixing angle lie in the lower octant. For models I and III, the effective Majorana mass $m_{\beta\beta}$ is within the reach of upcoming neutrinoless double-beta decay experiments, while it is too small to be detected in model II. Predicted leptonic {\it CP}-violating phases, and the sum of the active neutrino masses span wide but distinctive ranges, enabling future experiments to test and differentiate the proposed models.
\end{abstract}

\maketitle
\section{Introduction}
\label{sec:intro}
Explaining the masses and mixing of fermions is one of the toughest puzzles in particle physics. In particular, the distinct mass hierarchies and mixing patterns observed in the quark and lepton sectors strongly indicate the need for physics beyond the standard model (SM). On the other hand, the discovery of neutrino oscillations \cite{Super, SNO} has highlighted the necessity for new mechanisms to explain neutrino masses, along with other fundamental properties such as their precise mass values and their nature—whether they are Dirac or Majorana particles—further suggesting the need for extensions of the SM. Moreover, unlike the quark sector where mixing angles are small and {\it CP} violation is well established, neutrino oscillation experiments have revealed that two leptonic mixing angles of the Pontecorvo-Maki-Nakagawa-Sakata (PMNS) lepton flavor mixing matrix are large \cite{PDG}, and hint at a potential {\it CP} violation in the lepton sector, although the latter remains unconfirmed. These results have motivated the use of different approaches to address the flavor puzzle, with finite non-Abelian discrete symmetries emerging as an attractive scheme for flavor model building \cite{Ishi,Altarelli,King:2013eh,King:2017guk}. In this framework, fields transform linearly under irreducible representations of these symmetries, providing a structured way to produce fermion mixing. This is achieved through the introduction of gauge-singlet scalar fields called flavons whose vacuum expectation values (VEVs) align in specific directions breaking spontaneously the flavor symmetry and eventually dictating the mixing of fermions and the structure of their mass matrices. However, since these symmetries cannot remain exact, an intricate construction of the flavon potential is required where the introduction of additional cyclic groups is often necessary to eliminate undesirable operators and ensure the correct vacuum alignment. This complexity in both the symmetry breaking and flavon potential construction renders these models challenging to manage and limits their simplicity and predictivity.

\bigskip

Modular invariance has recently emerged as a promising approach to address the flavor puzzle, offering a simpler alternative to traditional flavor symmetries by avoiding the complexities of flavon vacuum alignment \cite{Feruglio}. This concept was first noted over three decades ago in certain types of string compactifications \cite{Ferrara:1989bc,Chun:1989se,Lauer:1990tm}, where Yukawa couplings are modeled as modular forms, which are holomorphic functions of a complex scalar field $\tau$, called the modulus. Motivated by the observed large neutrino mixing, this idea has been introduced by Feruglio in a bottom-up approach to construct explicit lepton mass models in the context of supersymmetry (SUSY) \cite{Feruglio}. In this scenario, matter fields and modular forms transform in irreducible representations under finite discrete groups $\Gamma_N$, which emerge as the quotient group of the modular group $\Bar{\Gamma}$ over the principal congruence subgroups $\Gamma(N)$ where $N$ is the level associated with each flavor group. An intriguing aspect of this scenario is that for $N \leq 5$, these finite flavor groups coincide with the known permutation groups, such as $S_3$, $A_4$, and $S_4$ \cite{B1}; see also Refs. \cite{Kobayashi:2023zzc, Ding:2023htn} for recent reviews on modular flavor symmetry, including lists of models implementing modular symmetries in the references therein. For a minimal setup, the modulus $\tau$ acquires a VEV at a high energy scale, serving as the only source of modular symmetry breaking. This disregards the need for additional flavons and simplifies the flavor structure by reducing the number of parameters required to fit the quark and lepton observables. A unified analysis of these observables necessitates a framework such as grand unified theories (GUTs) \cite{Pati,Pati:1973rp,Georgi:1974sy,Georgi:1974yf,Georgi:1974my,Fritzsch:1974nn}, where imposing a flavor symmetry alongside GUTs has been shown to address the flavor puzzle effectively \cite{King:2017guk}. When modular symmetries are combined with GUTs, the connection becomes more direct, as the modulus $\tau$ is shared between both quarks and leptons.

\bigskip

Among the known GUTs, the Pati-Salam (PS) model, based on the gauge group $SU(4)_C \times SU(2)_L \times SU(2)_R$, unifies quarks and leptons of the same $SU(2)$ isospin and provides a compelling framework to explaining phenomena beyond the SM \cite{Pati, Pati:1973rp}. For example, the presence of $SU(2)_R$ gauge symmetry predicts the existence of three right-handed (RH) neutrino states. Therefore, neutrino mass can be easily incorporated through the seesaw mechanism. Additionally, in contrast to the SM where each generation of fermions is organized into six multiplets (including RH neutrinos), the PS model unifies quarks and leptons into only two multiplets per generation. The minimal\footnote{The minimal Higgs sector with the smallest number of scalars required to generate fermion mass matrices.} version of this unification results in shared Yukawa couplings, leading to the prediction of mass matrix equalities: $M_e=M_d$ for charged leptons and down quarks, and $M_D=M_u$ for Dirac neutrinos and up quarks. To break these equalities, extension of the scalar sector by introducing additional Higgs multiplets is necessary. In this extended framework, Clebsch-Gordan (CG) coefficients are introduced into the Yukawa couplings allowing for distinct mass matrices for the different fermion sectors. As a partially unified model, the PS gauge group arises naturally as a subgroup of larger GUT frameworks, such as $SO(10)$, where it serves as an important intermediate step in the symmetry breaking chain of $SO(10)$ down to the SM gauge group \cite{Aulakh:2000sn}. Notably, unlike $SO(10)$ and other GUTs which are susceptible to rapid proton decay mediated by gauge bosons, the PS model inherently avoids this issue \cite{Davidson:1978pm, Mohapatra:1980qe}. These features allow the PS model to survive at relatively low energy scales \cite{Dolan:2020doe}, making it an attractive candidate for phenomenological studies.

\bigskip

In this paper, we built the first $\Gamma_2 \cong S_3$ flavor model with modular invariance in the framework of supersymmetric PS GUT, and analyze its predictions for fermion masses and mixings in both the quark and lepton sectors. As far as we are aware, this work constitutes only the second attempt to incorporate modular invariance into PS unification\footnote{While several PS models have been developed based on conventional non-Abelian discrete flavor symmetries \cite{Toor,King:2006np,King:2013hoa,King:2014iia,CarcamoHernandez:2017owh}, no PS model incorporating the standard $S_3$ group has been presented in the literature. However, a PS model extended by $S_{3L} \times S_{3R}$ chiral flavor symmetry has been proposed to realize democratic Yukawa matrices \cite{Yang:2016crz}.}. The first such implementation, presented in Ref. \cite{Ding}, explored several models based on the transformation properties of the modular $A_4$ group. These models have led to significant phenomenological predictions regarding fermion masses and their mixing. A modular $S_3$ group was first employed in supersymmetric models in Ref. \cite{Meloni:2023aru} to study lepton masses and mixings, and later in Ref. \cite{Marciano:2024nwm} to account for large low-energy \textit{CP} violation in the lepton sector and the matter-antimatter asymmetry of the Universe. In the context of GUTs, its application has so far been limited to the $SU(5)$ model \cite{Kobayashi:2019rzp,Du:2020ylx}. Additionally, neutrino phenomenology was explored in Ref. \cite{Behera:2024ark} for different realization of the $S_3$ modular symmetry based on the Type I seesaw mechanism, while Ref. \cite{Nomura:2024ouj} utilized $S_3$ modular invariance to study the leptonic dipole operator in relation to the anomalous magnetic moment of the muon. In our modular $S_3$-based Pati-Salam model, the matter multiplets $F_i \sim (4,2,1)$ and $F_i^c \sim (\Bar{4},1,2)$ can belong to either a doublet or the two singlets of the $S_3$ modular group while the Yukawa couplings are transformed as modular forms under the $S_3$ symmetry. Here, we propose three benchmark renormalizable models distinguished by the $S_3$ transformations of $F$ and $F_i^c$ and their modular weights. 
A vast number of models can, in principle, be constructed using various $S_3$ transformations and modular weight assignments for the fields. To narrow down this diversity, we have opted to use the same set of scalar fields across our three benchmark models, ensuring that their $S_3$ representations and modular weight assignments remain identical in all cases. We perform a numerical analysis to determine the values of free parameters, including the modulus $\tau$, that provide a consistent fit to the experimental observables in the lepton and quark sectors for each model. Our results show that all models prefer the normal mass ordering (NO) over the inverted ordering (IO), with the modulus $\tau$ constrained to narrow regions within the fundamental domain. Additionally, we present predictions for the neutrino mass parameters $\sum m_i$, $m_\beta$, $m_{\beta\beta}$, and the {\it CP}-violating phases. In the NO case, all models prefer the lower octant for the atmospheric angle, and the neutrino mass parameters are expected to be within the reach of future experimental sensitivities. In the quark sector, we find that the predicted mass ratios and mixing angles agree with experimental data at the GUT scale, except for a slight discrepancy in model I, where one parameter falls outside its $3\sigma$ range.

\bigskip 

The paper is structured as follows: Section \ref{sec2} reviews the scalar and fermion sectors of PS unification and revisits the derivation of fermion mass matrices. Section \ref{sec3} provides a brief overview of modular invariance and modular forms of level $N = 2$. In Section \ref{sec4}, we present three SUSY PS flavor models based on the finite modular group $\Gamma_2 \cong S_3$, detailing the fermion Yukawa matrices for each model. Section \ref{sec5} features a numerical analysis of these benchmark models for both NO and IO neutrino mass spectra, including predictions for fermion mass ratios, mixing parameters, and neutrino mass-related observables such as $m_{\beta\beta}$, $m_\beta$, and $\sum m_i$. Finally, Section \ref{concl} summarizes the findings and discusses the PS gauge symmetry breaking scale for the three models. Additional details on the $S_3$ modular group and higher-weight modular forms of level 2 are provided in the appendix.
\section{Revisiting the Pati-Salam unification}
\label{sec2}
The PS GUT unifies the quarks and leptons of a given chirality for each family into two representations of the PS gauge group \cite{Pati}. The latter extends the SM gauge group as $G_{PS} = SU(4)_C \times SU(2)_L \times SU(2)_R$ where the strong interaction symmetry $SU(3)_C$ is expanded to $SU(4)_C$, and $SU(2)_R$ acts as the right-handed counterpart to the familiar $SU(2)_L$ weak interaction. Thus, the SM matter fields and the RH neutrino of each generation are embedded in the following two chiral multiplets
\begin{equation}
    F_i \sim (4,2,1) = \begin{pmatrix}
u_i^r & u_i^g & u_i^b & \nu_i \\
d_i^r & d_i^g & d_i^b & e_i
\end{pmatrix} , \quad F_i^c \sim (\Bar{4},1,2) = \begin{pmatrix}
d_i^{rc} & d_i^{gc} & d_i^{bc} & e_i^c \\
-u_i^{rc} & -u_i^{gc} & -u_i^{bc} & -\nu_i^c
\end{pmatrix}
\end{equation}
where $F_i$ denotes the left-handed fermion multiplet, $F^c$ is the {\it CP} conjugate of the right-handed fermion multiplet while $i=1,2,3$ denotes the family index, and the superscripts $(r,g,b)$ are color indices. The decomposition of these matter multiplets under the SM gauge group is given by
\begin{eqnarray}
    F_i &\xrightarrow{G_{SM}}& (3,2,1/6) + (1,2,-1/2) \equiv Q_i + L_i \nonumber \\ F_i^c &\xrightarrow{G_{SM}}& (\Bar{3},1,-2/3) + (\Bar{3},1,1/3) + (1,1,1) + (1,1,0)\equiv u_i^c + d_i^c + e_i^c + \nu_i^c,
\end{eqnarray}
where $Q_i=(u_{iL},d_{iL})^T$ and $L_i=(\nu_{iL},l_{iL})^T$ stand for the three generations of left-handed quark and lepton fields, respectively. The $U(1)$ hypercharge generator $Y$ is a linear combination between the diagonal generator of $SU(2)_R$ and the $SU(4)_C$ generator of $B-L$
\begin{eqnarray}
    Y = I_{R_3} + \frac{B-L}{2}
\end{eqnarray}
For the scalar sector, the minimal set of Higgs multiplets required to realize successful symmetry breaking of the PS group down to $SU(3)_C \otimes U(1)_{em}$, and at the same time reproduce realistic fermion masses and mixing, can be identified by examining the decomposition of the tensor product of the fermion bilinears 
\begin{equation}
(4,2,1) \otimes (\Bar{4},1,2) = (1 \oplus 15, 2, 2) \quad \text{and} \quad (\Bar{4},1,2) \otimes (\Bar{4},1,2) = (\Bar{6} \oplus \Bar{10}, 1, 1 \oplus 3).
\label{tpro}
\end{equation}
From the first decomposition, it has been well established that at least two scalar multiplets, denoted as $\Phi$ and $\Sigma$, are required to generate viable masses for charged fermions. These scalars transform under the PS symmetry group $G_{PS}$ as $\Phi = (1, 2, 2)$ and $\Sigma = (15, 2, 2)$. A key distinction between these fields is that $\Phi$ is a color singlet under $SU(4)_C$, while $\Sigma$ carries color charges. If only $\Phi$ is present in the model, the resulting fermion mass relations are
\begin{eqnarray}
    m_e = m_d, \quad m_\mu = m_s, \quad m_\tau = m_b,
    \label{wmr}
\end{eqnarray} 
implying identical ratios for $m_e/m_\mu$ and $m_d/m_s$, which conflicts with experimental observations at both low and GUT energy scales \cite{PDG,Ross}. The inclusion of $\Sigma$ with an appropriately aligned VEV, modifies these problematic mass relations as will be discussed later. The second fermion bilinear in Eq.~\ref{tpro} coupled to a scalar multiplet transforming as $\Delta_R=(10, 1, 3)$ under $G_{PS}$ leads to the generation of Majorana masses for the RH neutrinos through the type-I seesaw mechanism \cite{Min,Yan,Gell,Glas,Moha}. On the other hand, in the framework of a left-right symmetric PS model, a left-handed $SU(2)_L$ Higgs triplet transforming as $\Delta_L=(\Bar{10}, 3, 1)$ under $G_{PS}$ is required for the preservation of the parity symmetry where $F \leftrightarrow F^c$ and $\Delta_R \leftrightarrow \Delta_L$. When $\Delta_L$ acquires a VEV, $\langle \Delta_L \rangle$, an additional contribution to neutrino masses arises through the type II seesaw mechanism \cite{Magg,Sche,Laz,Moha2}. \newline
The decomposition of the above scalar multiplets under the SM gauge group is given by 
\begin{eqnarray}
    \label{deco}
    \Phi &\xrightarrow{G_{SM}}& (1,2,1/2) + (1,2,-1/2) \equiv \Phi_u + \Phi_d \nonumber  \\
    \Sigma &\xrightarrow{G_{SM}}& (1,2,1/2) + (1,2,-1/2) + (3,2,1/6) + (\Bar{3},2,-1/6) + (3,2,7/6) \nonumber \\ &+& (\Bar{3},2,-7/6) + (8,2,1/2) + (8,2,-1/2) \equiv \Sigma_u + \Sigma_d + \Sigma_3 + \Sigma_{\Bar{3}} + \Sigma_4 + \Sigma_{\Bar{4}} + \Sigma_8 + \Sigma_{\Bar{8}}  \\
    \Delta_R &\xrightarrow{G_{SM}}& (1,1,0) + (1,1,-1) + (1,1,-2) + (3,1,2/3) + (3,1,-1/3)  + (3,1,-4/3) \nonumber \\ &+& (6,1,4/3) + (6,1,-1/3) + (6,1,-2/3) \equiv \Delta_0 + \Bar{\Delta_0} + \Tilde{\Delta_0} + \Delta_3 + \Delta_{\Bar{3}} + \Delta_4 + \Delta_6 + \Delta_{\Bar{6}} + \Delta_7 \nonumber
\end{eqnarray}
The breaking of the PS symmetry to the SM gauge group can proceed through multiple steps, depending on the scales at which the Higgs multiplets in the model acquire their VEVs. As shown in Eq. \ref{deco}, the multiplet $\Delta_R$ includes a SM singlet direction given by $\Delta_0 = (1,1,0)$, allowing for a single-step breaking to $G_{SM}$ when this singlet acquires a VEV $\langle \Delta_0 \rangle \sim \upsilon_R$. This breaking also induces RH Majorana neutrino masses proportional to $\upsilon_R$; the scale associated with this symmetry breaking. A two-step symmetry breaking can be achieved by introducing additional Higgs multiplets, specifically $\phi = (15, 1, 1)$ and $\Bar{\Delta}_R = (\Bar{10}, 1, 3)$. A non-zero VEV for $\phi$ breaks $SU(4)_C$ down to one of its maximal subgroups, $SU(3)_C \times U(1)_{B-L}$, giving rise to the well-known left-right symmetric group, $G_{LR} = SU(3)_C \times SU(2)_L \times SU(2)_R \times U(1)_{B-L}$. Subsequently, the VEVs of $\Delta_R \oplus \Bar{\Delta}_R$ break $SU(2)_R \times U(1)_{B-L}$ down to the SM hypercharge group $U(1)_Y$, resulting in the SM gauge group\footnote{See also Ref. \cite{Hart} for a comprehensive overview of possible breaking chains consistent with gauge-coupling unification, excluding the $\Delta_R \oplus \Bar{\Delta}_R$ multiplets.} $G_{SM}$ \cite{Melfo, Toor}. This two-step breaking pattern can be summarized as follows
\begin{eqnarray}
    SU(4)_C \times SU(2)_L \times SU(2)_R \xrightarrow{\langle \phi \rangle} SU(3)_C \times SU(2)_L \times SU(2)_R \times U(1)_{B-L} \xrightarrow{\langle \Delta_R \rangle} SU(3)_C \times SU(2)_L \times U(1)_Y
\end{eqnarray}
Finally, as shown in the decomposition of the fields $\Phi$ and $\Sigma$ in Eq. \ref{deco}, each of these fields contains two $SU(2)_L$ doublets that acquire non-zero VEVs, completing the symmetry breaking chain from the PS gauge group down to $SU(3)_C \times U(1)_Q$ and generating the charged fermion masses. To this end, the most general renormalizable Yukawa superpotential in the PS GUT is given by
\begin{eqnarray}
W_Y = F_i^c (Y_{ij}^1 \Phi + Y_{ij}^{15} \Sigma) F_j + Y_{ij}^{10_R} F_i^c F_j^c \Delta_R,
\label{wy}
\end{eqnarray}
where $Y_{ij}^1$ and $Y_{ij}^{15}$ are generally complex $3 \times 3$ matrices in family space, and $Y_{ij}^{10_R}$ is a symmetric $3 \times 3$ matrix. In the representation $F_i = (Q_i^r, Q_i^g, Q_i^b, L_i)$, the SM gauge group is broken by the VEVs of the Higgs multiplets $\Phi$ and $\Sigma$, given by
\begin{eqnarray}
    \langle \Phi \rangle = \text{diag}(1, 1, 1, 1) \times \begin{pmatrix} 
        \upsilon_\Phi^u & 0 \\ 
        0 & \upsilon_\Phi^d 
    \end{pmatrix}, \quad 
    \langle \Sigma \rangle = \text{diag}(1, 1, 1, -3) \times \begin{pmatrix} 
        \upsilon_\Sigma^u & 0 \\ 
        0 & \upsilon_\Sigma^d 
    \end{pmatrix}.  
    \label{VEVS}
\end{eqnarray}
As previously mentioned, the field $\Phi$ alone results in the wrong mass relations in Eq. \ref{wmr} because it is color blind and thus its VEV $\langle \Phi \rangle$ conserves the symmetry between quarks and leptons. This is why we introduce the Higgs field $\Sigma$ which transforms in the adjoint representation of $SU(4)_C$ and thus is represented by a traceless Hermitian matrix. Therefore, to preserve $SU(3)_C \otimes U(1)_Q$,  $\langle \Sigma \rangle$ must align in the specified direction of Eq. \ref{VEVS}. This introduces an additional CG factor of $-3$ for the leptons, which helps reconcile the mass spectrum differences between down-type quarks and charged leptons, ultimately leading to the Georgi–Jarlskog mass relations at the GUT scale \cite{GJ}. Based on the superpotential in Eq. \ref{wy} and the VEVs of $\Phi$ and $\Sigma$ in Eq. \ref{VEVS}, we derive the fermion mass matrices for the up-type quarks, down-type quarks, charged leptons, and Dirac neutrinos after electroweak (EW) symmetry breaking
\begin{eqnarray}
    M_d = Y^1 \upsilon_\Phi^d + Y^{15} \upsilon_\Sigma^d, \quad M_u = Y^1 \upsilon_\Phi^u + Y^{15} \upsilon_\Sigma^u, \nonumber \\ M_e = Y^1 \upsilon_\Phi^d - 3 Y^{15} \upsilon_\Sigma^d, \quad M_D = Y^1 \upsilon_\Phi^u - 3 Y^{15} \upsilon_\Sigma^u
    \label{duev}
\end{eqnarray}
Prior to this, the PS gauge symmetry is broken by the VEV of the Higgs multiplet $\Delta_R$, which induces a Majorana mass for the right-handed neutrinos given by $M_R = Y^{10_R}\langle \Delta_R \rangle$, where
\begin{eqnarray}
    \langle \Delta_R \rangle = \text{diag}(0,0,0,1) \times \begin{pmatrix}
        0 & 0 \\
        \upsilon_R & 0
    \end{pmatrix}
\end{eqnarray}
with $\langle \Delta_R \rangle \gg \langle \Phi \rangle \sim \langle \Sigma \rangle$. In this framework, the light neutrino masses emerge through the type-I seesaw mechanism, with the light neutrino mass matrix expressed as $m_\nu = - M_D^T M_R^{-1} M_D$, where the Dirac mass matrix $M_D$ is defined as in Eq. \ref{duev}. To simplify the numerical analysis following the diagonalization of the mass matrices, it is useful to redefine the Yukawa mass matrices in \ref{duev} in terms of the VEVs of the MSSM Higgs doublets, $\langle H_u \rangle = \upsilon_u$ and $\langle H_d \rangle = \upsilon_d$ \cite{Ding}
\begin{eqnarray}
   \Tilde{Y}^1 = \frac{\upsilon_\Phi^u}{\upsilon_u} Y^1, \quad \Tilde{Y}^{15} = \frac{\upsilon_\Sigma^d}{\upsilon_u} \frac{\upsilon_\Phi^u}{\upsilon_\Phi^d} Y^{15}, \quad r_1 = \frac{\upsilon_\Phi^d}{\upsilon_\Phi^u} \frac{\upsilon_u}{\upsilon_d}, \quad r_2 = \frac{\upsilon_\Sigma^u}{\upsilon_\Sigma^d} \frac{\upsilon_\Phi^d}{\upsilon_\Phi^u} 
\end{eqnarray}
where the mixing parameters $r_1$ and $r_2$ relate the VEVs of the MSSM Higgs doublets to those of the PS Higgs multiplets $\Phi$ and $\Sigma$. The redefined Yukawa matrices $\Tilde{Y}^1$ and $\Tilde{Y}^{15}$ are proportional to the original matrices $Y^1$ and $Y^{15}$, respectively, and the coefficients $\frac{\upsilon_\Phi^u}{\upsilon_u}$ and $\frac{\upsilon_\Sigma^d}{\upsilon_u} \frac{\upsilon_\Phi^u}{\upsilon_\Phi^d} Y^{15}$ can be absorbed into the coupling constants of each matrix. As a result, the mass matrices in Eq. \ref{duev} can be re-expressed in terms of $Y^1$, $Y^{15}$, $r_{1,2}$ and $\upsilon_{u,d}$ as follows
\begin{eqnarray}
 M_d = r_1( \Tilde{Y}^1 + \Tilde{Y}^{15}) \upsilon_d, \quad M_u = ( \Tilde{Y}^1 + r_2 \Tilde{Y}^{15}) \upsilon_u, \nonumber \\ M_e = r_1( \Tilde{Y}^1 - 3 \Tilde{Y}^{15}) \upsilon_d, \quad M_D = ( \Tilde{Y}^1 - 3 r_2 \Tilde{Y}^{15}) \upsilon_u 
 \label{duev2}
\end{eqnarray}
Notice that the mass matrices for the down-type quarks and charged leptons share the same parameter set and an overall factor $r_1 \upsilon_d$, differing only by the CG factor $-3$. Likewise, the up-type quark and Dirac mass matrices are parametrized similarly, with a common factor $\upsilon_u$, and differ solely due to the CG factor $-3$.
\section{Modular group of level $N=2$}
\label{sec3}
Modular symmetry is crucial in constructing supersymmetric theories, providing a more structured framework than traditional non-Abelian discrete symmetries widely used in the last two decades to describe fermion masses and mixing. The homogeneous modular group, denoted as $\Gamma \equiv SL(2, \mathbb{Z})$, consists of $2 \times 2$ matrices with integer coefficients and unit determinant, and serves as the foundation for modular symmetry studies. For model building purposes, we typically use the inhomogeneous modular group, also known as the projective special linear group, which is the quotient of the two-dimensional special linear group by its center: $\Bar{\Gamma} = PSL(2, \mathbb{Z}) \equiv SL(2, \mathbb{Z})/\{I_2, -I_2\}$, where $I_2$ is the two-dimensional identity matrix. This group consists of linear fractional transformations, denoted by $\gamma$, acting on the modulus $\tau$ in the upper-half complex plane $H = \{\tau \in \mathbb{C} \mid \text{Im}(\tau) > 0\}$, defined as
\begin{equation}
\gamma(\tau) = \frac{a\tau + b}{c\tau + d}, \quad a, b, c, d \in \mathbb{Z}, \quad ad - bc = 1. 
\end{equation}
These transformations can be generated by two fundamental elements, $S$ and $T$, which act on $\tau$ as follows
\begin{equation}
S: \tau \to -\frac{1}{\tau}, \quad T: \tau \to \tau + 1.  
\label{th2}
\end{equation}
These generators satisfy the relations $S^2 = (ST)^3 = 1$, and from the form of the transformations in Eq. \ref{th2}, their representation matrices are given by
\begin{equation}
S = \begin{pmatrix} 0 & 1 \\ -1 & 0 \end{pmatrix}, \quad T = \begin{pmatrix} 1 & 1 \\ 0 & 1 \end{pmatrix}.    
\end{equation}
For $T^N=1$, a family of $PSL(2, \mathbb{Z})$ normal subgroups denoted as $\bar{\Gamma}(N)$ is introduced, where $N$ represents the level of the group. These infinite subgroups, called congruence subgroups, act on $\tau$ similarly to $\Gamma$ but with additional modular congruence conditions that constrain the values of $a$, $b$, $c$, and $d$ modulo $N$
\begin{equation}
    \bar{\Gamma}(N) = \left\{ \begin{pmatrix} a & b \\ c & d \end{pmatrix} \in SL(2, \mathbb{Z}) \Bigg| \begin{pmatrix} a & b \\ c & d \end{pmatrix} \equiv \begin{pmatrix} 1 & 0 \\ 0 & 1 \end{pmatrix} \mod N \right\}.
\end{equation}
For $N=2$, $\bar{\Gamma}(2)$ is defined as the quotient $\Gamma(2)/\{I, -I\}$, while for $N>2$, $\bar{\Gamma}(N) \equiv \Gamma(N)$ because $-I$ is not an element of $\Gamma(N)$. The quotient groups $\Gamma_N \equiv \bar{\Gamma}/\bar{\Gamma}(N)$ are known as finite modular groups. These groups are particularly interesting for $N \leq 5$, as they are isomorphic to the well-known permutation groups that are widely used in building flavor models \cite{B1}. Specifically, $\Gamma_2 \simeq S_3$, $\Gamma_3 \simeq A_4$, $\Gamma_4 \simeq S_4$, and $\Gamma_5 \simeq A_5$. The main difference between ordinary non-Abelian flavor groups and modular flavor groups is that, in the former, the transformation properties are purely group-theoretic, with the invariance of the superpotential determined solely by the structure of the discrete group, while the coupling constants do not transform under the action of the flavor group. In contrast, modular flavor symmetries require modular invariance where the theory remains unchanged under modular group transformations, and the Yukawa couplings transform non-trivially as modular forms--functions of the modulus $\tau$. These modular forms are holomorphic functions characterized by a positive integer level $N$ and a non-negative integer weight $k$, and they satisfy the following condition
\begin{equation}
f(\gamma(\tau)) = (c\tau + d)^k f(\tau).    
\end{equation}
Moreover, the modular forms of a given weight $k$ and level $N$ form a finite-dimensional linear space, denoted as $M_k(\Gamma(N))$. For even weights, a basis can be chosen such that the modular forms transform under a unitary irreducible representation $\rho$ of $\Gamma_N$. Then, when modular forms are decomposed into representations of the modular group, they often come in multiplets $f_{\bm{r}}(\tau) = (f_1(\tau), f_2(\tau), f_3(\tau), ...)^T$ which transform as \cite{Feruglio}
\begin{eqnarray}
    f_{\bm{r}}(\gamma \tau) = (c\tau + d)^k \rho_{\bm{r}}(\gamma) f_{\bm{r}}(\tau) , \quad \gamma \in \Gamma_N
    \label{frg}
\end{eqnarray}
where $\rho_{\bm{r}}(\gamma)$ is the representation matrix of $\gamma$ in the irreducible representation $\bm{r}$. \newline
In a modular invariant supersymmetric theory, a superpotential $\mathcal{W}(\Psi)$ is a holomorphic function of the chiral superfields $\Psi$ which depend on the modulus $\tau$ and chiral supermultiplets $\psi^I$, with $\Psi = (\tau, \psi^I)$. These supermultiplets $\psi^I$ transform under the modular group as $\psi^I \rightarrow (c\tau + d)^{-k_I} \rho^I \psi^I$ and they are not modular forms while the value of $k_I$ is unrestricted. Consider the following expansion of $\mathcal{W}(\tau,\psi^I)$ in terms of the chiral supermultiplets $\psi^I$
\begin{eqnarray}
    \mathcal{W}(\tau,\psi^I) = \sum_n Y_{I_1...I_n}(\tau) \psi^{I_1}...\psi^{I_n}
\end{eqnarray}
In order to ensure that each term in $\mathcal{W}(\tau,\psi^I)$ is modular invariant, the modular forms $Y_{I_1...I_n}(\tau)$ must transform according to Eq. \ref{frg} as
\begin{eqnarray}
    Y_{I_1...I_n}(\tau) \rightarrow Y_{I_1...I_n}(\gamma \tau) =  (c\tau + d)^{k_Y} \rho_{\bm{r}_Y}(\gamma) Y_{I_1...I_n}(\tau)
\end{eqnarray}
where $k_Y = k_{I_1} + \dots + k_{I_n}$ compensates the weights of $\psi^{I_1}...\psi^{I_n}$, and the product $\rho_{\bm{r}_Y} \otimes \rho^{I_1} \otimes ... \otimes \rho^{I_n}$ contains an invariant singlet of $\Gamma_N$; $\rho_{\bm{r}_Y} \otimes \rho^{I_1} \otimes ... \otimes \rho^{I_n} \supset 1$.

In this work, we are interested in the symmetry group of level $N=2$ which is isomorphic to the permutation group of three objects $S_3$. This group consists of three irreducible representations: a doublet $\mathbf{2}$, a trivial singlet $\mathbf{1}$, and a pseudo-singlet $\mathbf{1}'$. The structure of $S_3$ allows for compact representations of modular forms that can be used in constructing flavor models. For $k = 2$, the smallest non-trivial modular forms can be represented by two independent functions $Y_1(\tau)$ and $Y_2(\tau)$ which form an irreducible doublet of $S_3$. These are typically generated by utilizing the Dedekind eta function $\eta(\tau)$, defined as follows
\begin{eqnarray}
\eta(\tau) = q^{1/24} \prod_{n=1}^{\infty} (1 - q^n), \quad q = e^{2\pi i \tau},
\end{eqnarray}
where $\eta(\tau)$ transforms under the generators of the modular group, $S$ and $T$, as follows
\begin{eqnarray}
\eta\left(-\frac{1}{\tau}\right) = \sqrt{-i\tau} \eta(\tau), \quad \eta(\tau + 1) = e^{i\pi/12} \eta(\tau).
\end{eqnarray}
The modular forms for $\Gamma_2 \equiv S_3$ can be expressed in terms of derivatives of the Dedekind eta function as follows
\begin{eqnarray}
Y_1(\tau) &=& \frac{1}{2} \left( \frac{\eta'(\tau/2)}{\eta(\tau/2)} + \frac{\eta'((\tau+1)/2)}{\eta((\tau+1)/2)} - 8 \frac{\eta'(2\tau)}{\eta(2\tau)} \right), \nonumber \\
Y_2(\tau) &=& \frac{\sqrt{3}}{2} \left( \frac{\eta'(\tau/2)}{\eta(\tau/2)} - \frac{\eta'((\tau+1)/2)}{\eta((\tau+1)/2)} \right),
\end{eqnarray}
where $\eta'(\tau)$ represents the derivative of the eta function with respect to $\tau$. In order to simplify numerical calculations, we implement the q-expansion form of these functions as shown below \cite{Koba} 
\begin{eqnarray}
    Y_2^{(2)} = \begin{pmatrix}
Y_1(\tau) \\ Y_2(\tau)
\end{pmatrix}_2 = \begin{pmatrix}
1/8 + 3q + 3q^2 + 12q^3 + 3q^4 + ... \\ \sqrt{3}q^{1/2} (1+4q+6q^2+8q^3+...)
\end{pmatrix}_2
\end{eqnarray}
These two functions form the foundation for constructing modular forms of higher weights which can be derived easily by taking tensor products of the lower-weight modular forms $Y_1(\tau)$ and $Y_2(\tau)$; see Appendix \ref{appe} for more details. 
\section{Implementing $\Gamma_2$ in Pati-Salam GUT}
\label{sec4}
We now turn to an analysis of the modular-invariant superpotential in various PS models depending on the weights and $S_3$ transformations of the fields in Eq. \ref{wy}. Generally, this renormalizable superpotential can be expressed as
\begin{eqnarray}
W_Y^{\Gamma_2} &=& \sum_{(R_x \otimes R_x^{\prime})_s} \left\{ \left[ a_s (F_i^c \Phi F_j)_{R_x}~Y_{R_x^{\prime}}^{(k_{F_i^c} + k_\Phi +k_{F_j})}(\tau) \right]_1 + \left[ b_s (F_i^c \Sigma F_j)_{R_x}~Y_{R_x^{\prime}}^{(k_{F_i^c} + k_\Sigma +k_{F_j})}(\tau) \right]_1 \right. \nonumber \\
&+&  \left. \left[ c_s (F_i^c F_j^c \Delta_R)_{R_x} ~Y_{R_x^{\prime}}^{(k_{F_i^c} + k_\Sigma +k_{F_j^c})}(\tau) \right]_1 \right\}\,
\end{eqnarray}
where $R_x \otimes R_x^{\prime}$ represent all tensor products that contain the $S_3$ trivial singlet (see Eq. \ref{A2} in appendix), while $s=1,2,3,...$ represent the coupling index for each of the $S_3$ invariant terms. There are numerous possible charge assignments, whether through $S_3$ irreducible representations or weight assignments, that can lead to a variety of models. To manage this complexity, we start with specific assumptions to limit the number of models and parameters, aiming to fit the model parameters to the experimental values of physical observables in both the lepton and quark sectors. For simplicity, we assume that the Higgs multiplets $\Phi$, $\Sigma$, and $\Delta_R$ transform trivially under the $S_3$ symmetry, and we limit ourselves to level-2 modular forms with weights up to 6. Moreover, to ensure distinct structures in the Yukawa matrices generated by the $\Phi$-associated terms $Y^1$ and the $\Sigma$-associated terms $Y^{10}$, we take $k_\Phi \neq k_\Sigma$. Finally, we do not consider the scenario where all fermions are assigned to $S_3$ singlets, as it effectively reduces to a simple $Z_2$ symmetry. Such a case not only fails to align with the primary motivation for employing modular flavor symmetries—namely, accommodating fermions in doublet and triplet representations for a more robust description of neutrino flavor mixing—but also requires the inclusion of additional free parameters, making the models more complex and the fit to physical observables more challenging.

\medskip
\textbf{Model I: $F_{1,2} \equiv \bm{2}$, $F_3 \equiv \bm{1^\prime}$, $F_{1,2}^c \equiv \bm{2}$, $F_3^c \equiv \bm{1^\prime}$}:
For our first model, we assign the first two generations of left-handed fermions, $F_{1,2}$, to the $S_3$ doublet $\bm{2}$, while the third generation, $F_3$, is assigned to an $S_3$ pseudo-singlet. The same assignment is used for the {\it CP} conjugate right-handed fermions: $F_{1,2}^c \sim \bm{2}$ and $F_3^c \sim \bm{1'}$.
\begin{table}[H]
\centering
\begin{tabular}{|c||c|c|c|c|c|c|c|c|c|c|}
\hline
Model I & $F_{1,2}$ & $F_3$ & $F_{1,2}^c$ & $F_3^c$ & $\Phi$ & $\Sigma$ & $\Delta_R$ & $Y_{\bm{2}}^{(2)}$ & $Y_{\bm{1}}^{(4)}$ & $Y_{\bm{2}}^{(4)}$  \\ \hline
$G_{PS}$ & $(4,2,1)$ & $(4,2,1)$ & $(\Bar{4},1,2)$ & $(\Bar{4},1,2)$ & $(1,2,2)$ & $(15,2,2)$ & $(10,1,3)$ & $1$ & $1$ & $1$ \\ \hline
$S_3$ & $\bm{2}$ & $\bm{1^\prime}$ & $\bm{2}$ & $\bm{1^\prime}$ & $\bm{1}$ & $\bm{1}$ & $\bm{1}$ & $\bm{2}$ & $\bm{1}$ & $\bm{2}$ \\ \hline
$k_I$ & $1$ & $1$ & $1$ & $1$ & $0$ & $2$ & $0$ & $2$ & $4$ & $4$ \\ \hline
\end{tabular}
\caption{The charge assignments of $G_{PS}$ $S_3$, and weight for the fields and modular forms used in model I.}
\label{tab1}
\end{table}
The $S_3$ irreducible representations and weight assignments for various fields, along with the modular forms required to ensure modular invariance of the superpotential in this model are presented in Table \ref{tab1}. The renormalizable Yukawa superpotential invariant under the PS gauge group and the $\Gamma_2$ modular group is given by
\begin{eqnarray}
    W_Y^I &=& a_1 F_{1,2}^c F_{1,2} \Phi Y_{\bm{2}}^{(2)} + a_2 F_{1,2}^c F_3 \Phi Y_{\bm{2}}^{(2)} + a_3 F_3^c F_{1,2} \Phi Y_{\bm{2}}^{(2)} + b_1 F_{1,2}^c F_{1,2} \Sigma Y_{\bm{1}}^{(4)} + b_2 F_{1,2}^c F_{1,2} \Sigma Y_{\bm{2}}^{(4)} + b_3 F_{1,2}^c F_3 \Sigma Y_{\bm{2}}^{(4)} \nonumber \\ &+& b_4 F_3^c F_{1,2} \Sigma Y_{\bm{2}}^{(4)} + b_5 F_3^c F_3 \Sigma Y_{\bm{1}}^{(4)} + c_1 F_{1,2}^c F_{1,2}^c \Delta_R Y_{\bm{2}}^{(2)} + c_2 F_{1,2}^c F_3^c \Delta_R Y_{\bm{2}}^{(2)}
\end{eqnarray}
Using the decomposition of the tensor product of $S_3$ irreducible representations in Eq. \ref{A2}, this superpotential leads to the following Yukawa matrices
\begin{eqnarray}
    Y^{15} &=& \begin{pmatrix}
        b_1 (Y_1^2 + Y_2^2) - b_2 (Y_2^2 - Y_1^2) & 2 b_2 Y_1 Y_2 & 2 b_3 Y_1 Y_2 \\
        2 b_2 Y_1 Y_2 & b_1 (Y_1^2 + Y_2^2) + b_2 (Y_2^2 - Y_1^2) & - b_3 (Y_2^2 - Y_1^2) \\
        2 b_4 Y_1 Y_2 & - b_4 (Y_2^2 - Y_1^2) & b_5 (Y_1^2 + Y_2^2)    
    \end{pmatrix}, \nonumber \\
    Y^1 &=& \begin{pmatrix}
        -a_1 Y_1 & a_1 Y_2 & a_2 Y_2 \\
        a_1 Y_2 & a_1 Y_1 & -a_2 Y_1 \\
        a_3 Y_2 & -a_3 Y_1 & 0 
    \end{pmatrix}, \quad \quad  
    Y^{10_R} = \begin{pmatrix}
        -c_1 Y_1 & c_1 Y_2 & c_2 Y_2 \\
        c_1 Y_2 & c_1 Y_1 & -c_2 Y_1 \\
        c_2 Y_2 & -c_2 Y_1 & 0
        \end{pmatrix}.
\label{yuk1}
\end{eqnarray}
To derive the fermion mass matrices, we follow the same procedure used to obtain their expressions in Eq. \ref{duev2}. Each Yukawa matrix is then factorized by a coupling constant, enabling us to express the input parameters for our numerical analysis as overall mass scales, coupling constant ratios, and the complex modulus $\tau$ associated with the modular forms. As an example, the charged lepton mass matrix is given by $M_e = r_1 (\Tilde{Y}^1 - 3 \Tilde{Y}^{15}) \upsilon_d$. Thus, following the above parametrization, $M_e$ is explicitly given by
\begin{eqnarray}
    M_e = r_1 a_1 \upsilon_d \left[
    \begin{pmatrix}
        -Y_1 & Y_2 & \frac{a_2}{a_1} Y_2 \\
        Y_2 & Y_1 & -\frac{a_2}{a_1} Y_1 \\
        \frac{a_3}{a_1} Y_2 & -\frac{a_3}{a_1} Y_1 & 0
    \end{pmatrix}
    -3
    \begin{pmatrix}
         \frac{b_1}{a_1} (Y_1^2 + Y_2^2) - \frac{b_2}{a_1} (Y_2^2 - Y_1^2) & 2 \frac{b_2}{a_1} Y_1 Y_2 & 2 \frac{b_3}{a_1} Y_1 Y_2 \\
        2 \frac{b_2}{a_1} Y_1 Y_2 & \frac{b_1}{a_1} (Y_1^2 + Y_2^2) + \frac{b_2}{a_1} (Y_2^2 - Y_1^2) & - \frac{b_3}{a_1} (Y_2^2 - Y_1^2) \\
        2 \frac{b_4}{a_1} Y_1 Y_2 & - \frac{b_4}{a_1} (Y_2^2 - Y_1^2) & \frac{b_5}{a_1} (Y_1^2 + Y_2^2)  
    \end{pmatrix} \right] . \nonumber
\end{eqnarray}
where $r_1 a_1 \upsilon_d$ serves as an overall scale factor, which also appears in the explicit expression for $M_d$. The same approach is applied to derive the other fermion mass matrices. The phases of the couplings $a_i$ associated with the $\Phi$-related terms, as well as $c_1$, can be absorbed through a redefinition of the matter fields, while all other couplings are complex. Consequently, the model includes a total of 21 independent real parameters, including the real and imaginary parts of the modulus $\tau$. 

\medskip
\textbf{Model II: $F_{1,2} \equiv \bm{2}$, $F_3 \equiv \bm{1}$, $F_{1}^c \equiv \bm{1}$, $F_{2}^c \equiv \bm{1^\prime}$, $F_3^c \equiv \bm{1}$}: In this configuration, the first two generations of left-handed fermions, $F_{1,2}$, are grouped into an $S_3$ doublet $\bm{2}$, while the third generation, $F_3$, is assigned to the $S_3$ trivial singlet. For the {\it CP}-conjugate right-handed fermions, the first and third generations transform as $S_3$ trivial singlets, whereas the second generation transforms as a pseudo-singlet under $S_3$: $F_1^c \sim \bm{1}$, $F_2^c \sim \bm{1^\prime}$, and $F_3^c \sim \bm{1}$. Table \ref{tab2} provides the $S_3$ and weight assignments for all fields, as well as the modular forms relevant to this model.
\begin{table}[H]
\centering
\begin{tabular}{|c||c|c|c|c|c|c|c|c|c|c|c|c|c|}
\hline
Model II & $F_{1,2}$ & $F_3$ & $F_1^c$ & $F_2^c$  & $F_3^c$ & $\Phi$ & $\Sigma$ & $\Delta_R$ & $Y_{\bm{2}}^{(2)}$ & $Y_{\bm{1}}^{(4)}$ & $Y_{\bm{2}}^{(4)}$ & $Y_{\bm{1}}^{(6)}$ & $Y_{\bm{2}}^{(6)}$ \\ \hline
$G_{PS}$ & $(4,2,1)$ & $(4,2,1)$ & $(\Bar{4},1,2)$ & $(\Bar{4},1,2)$ & $(\Bar{4},1,2)$ & $(1,2,2)$ & $(15,2,2)$ & $(10,1,3)$ & $1$ & $1$ & $1$  & $1$  & $1$  \\ \hline
$S_3$ & $\bm{2}$ & $\bm{1}$ & $\bm{1}$ & $\bm{1^\prime}$ & $\bm{1}$ & $\bm{1}$ & $\bm{1}$ & $\bm{1}$ & $\bm{2}$ & $\bm{1}$ & $\bm{2}$  & $\bm{1}$  & $\bm{2}$  \\ \hline
$k_I$ & $2$ & $2$ & $0$ & $0$ & $2$ & $0$ & $2$ & $0$ & $2$ & $4$ & $4$  & $6$  & $6$   \\ \hline
\end{tabular}
\caption{The charge assignments of $G_{PS}$, $S_3$ and weights for the fields and modular forms used in model II.}
\label{tab2}
\end{table}
The renormalizable Yukawa superpotential invariant under $G_{PS}$ and $\Gamma_2$ can be expressed as
\begin{eqnarray}
    W_Y^{II} &=& a_1 F_1^c F_{1,2} \Phi Y_{\bm{2}}^{(2)} + a_2 F_2^c F_{1,2} \Phi Y_{\bm{2}}^{(2)} + a_3 F_3^c F_{1,2} \Phi Y_{\bm{2}}^{(4)} + a_4 F_3^c F_3 \Phi Y_{\bm{1}}^{(4)} + b_1 F_1^c F_{1,2} \Sigma Y_{\bm{2}}^{(4)}  \nonumber \\ &+& b_2 F_1^c F_3 \Sigma Y_{\bm{1}}^{(4)} + b_3 F_2^c F_{1,2} \Sigma Y_{\bm{2}}^{(4)} + b_4 F_3^c F_{1,2} \Sigma Y_{\bm{2}}^{(6)} + b_5 F_3^c F_3 \Sigma Y_{\bm{1}}^{(6)} \\ &+& c_1 F_1^c F_1^c \Delta_R + c_2 F_2^c F_2^c \Delta_R + c_3 F_3^c F_3^c \Delta_R Y_{\bm{1}}^{(4)} \nonumber
\end{eqnarray}
By using the tensor product of $S_3$ irreducible representations given in Eq. \ref{A2}, we obtain the following Yukawa matrices
\begin{eqnarray}
    Y^1 &=& \begin{pmatrix}
        a_1 Y_1 & a_1 Y_2 & 0 \\
        a_2 Y_2 & -a_2 Y_1 & 0 \\
        a_3 (Y_2^2-Y_1^2) & 2 a_3 Y_1 Y_2 & a_4 (Y_1^2 + Y_2^2) 
    \end{pmatrix}, \quad \quad  
    Y^{10_R} = \begin{pmatrix}
        c_1 & 0 & 0 \\
        0 & c_2 & 0 \\
        0 & 0 & c_3 (Y_1^2 + Y_2^2)
        \end{pmatrix},  \nonumber \\
    Y^{15} &=& \begin{pmatrix}
        b_1 (Y_2^2 - Y_1^2) & 2 b_1 Y_1 Y_2 & b_2 (Y_1^2 + Y_2^2) \\
        2 b_3 Y_1 Y_2 & -b_3 (Y_2^2 - Y_1^2)  & 0 \\
        b_4 (Y_1 Y_2^2 + Y_1^3) & b_4 (Y_2^3 + Y_1^2 Y_2) & b_5 (3 Y_1 Y_2^2 - Y_1^3)    
    \end{pmatrix}.
\label{yuk2}
\end{eqnarray}
Similar to model I, the couplings $a_i$ and $c_1$ can be made real through a redefinition of the matter fields, while the remaining couplings are complex. Therefore, the model comprises a total of $24$ free parameters. 

\smallskip

\textbf{Model III: $F_1 \equiv \bm{1}$, $F_2 \equiv \bm{1}$, $F_3 \equiv \bm{1}$, $F_{1,2}^c \equiv \bm{2}$, $F_{3}^c \equiv \bm{1}$}: In this model, the three generations of left-handed fermions, $F_{i}$, transform trivially under $S_3$, while the {\it CP}-conjugate right-handed fermions are assigned as $\bm{2} + \bm{1}$ under $S_3$ with $F_{1,2}^c \sim \bm{2}$ and $F_3^c \sim \bm{1}$. Table \ref{tab3} provides the $S_3$ and weight assignments for all fields, as well as the modular forms relevant to this model.
\begin{table}[H]
\centering
\begin{tabular}{|c||c|c|c|c|c|c|c|c|c|c|c|c|}
\hline
Model III & $F_1$ & $F_2$ & $F_3$ & $F_{1,2}^c$  & $F_3^c$ & $\Phi$ & $\Sigma$ & $\Delta_R$ & $Y_{\bm{2}}^{(2)}$ & $Y_{\bm{1}}^{(4)}$ & $Y_{\bm{2}}^{(4)}$ & $Y_{\bm{2}}^{(6)}$ \\ \hline
$G_{PS}$ & $(4,2,1)$ & $(4,2,1)$ & $(\Bar{4},1,2)$ & $(\Bar{4},1,2)$ & $(\Bar{4},1,2)$ & $(1,2,2)$ & $(15,2,2)$ & $(10,1,3)$ & $1$ & $1$ & $1$  & $1$  \\ \hline
$S_3$ & $\bm{1}$ & $\bm{1}$ & $\bm{1}$ & $\bm{2}$ & $\bm{1}$ & $\bm{1}$ & $\bm{1}$ & $\bm{1}$ & $\bm{2}$ & $\bm{1}$ & $\bm{2}$   & $\bm{2}$  \\ \hline
$k_I$ & $0$ & $2$ & $2$ & $2$ & $0$ & $0$ & $2$ & $0$ & $2$ & $4$ & $4$ & $6$   \\ \hline
\end{tabular}
\caption{The charge assignments under $G_{PS}$, $S_3$ and weights for the fields and modular forms used in model III.}
\label{tab3}
\end{table}
The renormalizable Yukawa superpotential invariant under the PS gauge group and the $S_3$ modular group reads as
\begin{eqnarray}
    W_Y^{III} &=& a_1 F_{1,2}^c F_{1} \Phi Y_{\bm{2}}^{(2)} + a_2 F_{1,2}^c F_{2} \Phi Y_{\bm{2}}^{(4)} + a_3 F_{1,2}^c F_{3} \Phi Y_{\bm{2}}^{(4)} + a_4 F_3^c F_1 \Phi + b_1 F_{1,2}^c F_{1} \Sigma Y_{\bm{2}}^{(4)}  \nonumber \\ &+& b_2 F_{1,2}^c F_2 \Sigma Y_{\bm{2}}^{(6)} + b_3 F_{1,2}^c F_{3} \Sigma Y_{\bm{2}}^{(6)} + b_4 F_3^c F_{2} \Sigma Y_{\bm{1}}^{(4)} + b_5 F_3^c F_3 \Sigma Y_{\bm{1}}^{(4)} \\ &+& c_1 F_{1,2}^c F_{1,2}^c \Delta_R Y_{\bm{1}}^{(4)} + c_2 F_{1,2}^c F_{1,2}^c \Delta_R Y_{\bm{2}}^{(4)} + c_3 F_{1,2}^c F_3^c \Delta_R Y_{\bm{2}}^{(2)} + c_4 F_3^c F_3^c \Delta_R \nonumber
\end{eqnarray}
By applying the decomposition of the tensor product of 
$S_3$ representations given in Eq. \ref{A2}, we derive the Yukawa matrices for this model given as follows
\begin{eqnarray}
    Y^1 &=& \begin{pmatrix}
        a_1 Y_1 & a_2 (Y_2^2 - Y_1^2) & a_3 (Y_2^2 - Y_1^2) \\
        a_1 Y_2 & 2 a_2 Y_1 Y_2 & 2 a_3 Y_1 Y_2 \\
        a_4 & 0 & 0  
    \end{pmatrix}, \quad \quad
        Y^{15} = \begin{pmatrix}
        b_1 (Y_2^2 - Y_1^2) & b_2 (Y_1 Y_2^2 + Y_1^3) & b_3 (Y_1 Y_2^2 + Y_1^3) \\
        2 b_1 Y_1 Y_2 & b_2 (Y_2 Y_1^2 + Y_2^3)  & b_3 (Y_2 Y_1^2 + Y_2^3) \\
        0 & b_4 (Y_1^2 + Y_2^2) & b_5 (Y_1^2 + Y_2^2)    
    \end{pmatrix}. \nonumber \\
      Y^{10_R} &=& \begin{pmatrix}
        c_1 (Y_1^2 + Y_2^2) - c_2 (Y_2^2 - Y_1^2) & 2 c_2 Y_1 Y_2 & c_3 Y_1 \\
        2 c_2 Y_1 Y_2 & c_1 (Y_1^2 + Y_2^2) + c_2 (Y_2^2 - Y_1^2) & c_3 Y_2 \\
        c_3 Y_1 & c_3 Y_2 & c_4
        \end{pmatrix}
\label{yuk3}
\end{eqnarray}
Using the procedure from model I, we derive the fermion mass matrices and find $26$ free parameters in model III. In the following section, we conduct a thorough numerical study of the three proposed models, uncovering regions in the parameter space that align exceptionally well with experimental data.
\section{Numerical results}
\label{sec5}
In this section, we provide a comprehensive numerical study of the predictions for the three benchmark PS models introduced earlier. The distinguishing features of these models lie in the transformation behavior of the matter fields $F_i$ and $F_i^c$ under the $S_3$ group and the modular weight assignments for each field. In our analysis, we employed modular forms of level 2 with weights up to 6. The Yukawa matrices for the three models are explicitly presented in Eqs.~\ref{yuk1}, \ref{yuk2}, and \ref{yuk3}. Each model relies on a set of dimensionless input parameters, including the modulus $\tau$, coupling constant ratios, and overall mass scales in the up-type quark, charged lepton/down-type quark, and neutrino mass matrices. These parameters ultimately determine the fermion mass ratios, mixing angles, and the \textit{CP}-violating phases.
The analysis is performed at the GUT scale, where the quark and charged lepton masses, along with the CKM mixing angles, are taken from Ref.~\cite{Ross}, assuming $\tan\beta = 10$ and a SUSY breaking scale of $ M_{SUSY} = 500~GeV$. For neutrino oscillation parameters, the latest global fit from NuFIT v6.0, incorporating Super-Kamiokande atmospheric data \cite{nufit}, is used. The renormalization group (RG) running effects on neutrino parameters are known to be small for a hierarchical neutrino mass spectrum \cite{Chankowski:2001mx,Antusch:2005gp}. Since our benchmark models, as discussed below, favor and exhibit the NO mass spectrum, we expect the RG corrections to be negligible and thus have little impact on our results.
\begin{table}[H]
\centering
\renewcommand{\arraystretch}{1.5}
\setlength{\tabcolsep}{10pt}
\begin{tabular}{|c|c||c|c|c|}
\hline
Parameters & $\mu_i \pm 1\sigma$ \cite{Ross} & Parameters & $\mu_i \pm 1\sigma$ & $3\sigma$ ranges \\ \hline
$m_e/m_\mu$ & $0.0048 \pm 0.0002$ &  $\Delta m_{21}^2/10^{-5}eV^2$ & $7.49 \pm 0.19$ & $6.92 \rightarrow 8.05$ \\ \hline
$m_\mu/m_\tau$ & $0.059 \pm 0.002$ & $\Delta m_{31}^2/10^{-3}eV^2$(NO)  & $2.513_{-0.019}^{+0.021}$ & $2.451 \rightarrow 2.578$ \\ \hline
$m_u/m_c$ & $0.0027 \pm 0.0006$ & $\Delta m_{32}^2/10^{-3}eV^2$(IO) & $-2.484 \pm 0.020$ & $-2.547 \rightarrow -2.421$ \\ \hline
$m_c/m_t$ & $0.0025 \pm 0.0002$ & $\sin^2\theta_{12}^l$ & $0.308_{-0.011}^{+0.012}$ & $0.275 \rightarrow 0.345$ \\ \hline
$m_d/m_s$ & $0.051 \pm 0.007$ & $\sin^2\theta_{23}^l$(NO) & $0.470_{-0.013}^{+0.017}$ & $0.435 \rightarrow 0.585 $ \\ \hline
$m_s/m_b$ & $0.019 \pm 0.002$ & $\sin^2\theta_{23}^l$(IO) & $0.550_{-0.015}^{+0.012}$ & $0.440 \rightarrow 0.584$ \\ \hline
$\theta_{12}^q$ & $0.229 \pm 0.001$ & $\sin^2\theta_{13}^l$(NO) & $0.02215_{-0.00058}^{+0.00056}$ & $0.02030 \rightarrow 0.02388$ \\ \hline
$\theta_{13}^q$ & $0.0037 \pm 0.0004$ & $\sin^2\theta_{13}^l$(IO) & $0.02231 \pm 0.00056$ & $0.02060 \rightarrow 0.02409$ \\ \hline
$\theta_{23}^q$ & $0.0397 \pm 0.0011$ & $\delta_{CP}^l/^\circ$(NO) & $212_{-41}^{+26}$ & $124 \rightarrow 364$ \\ \hline
$\delta_{CP}^q/^\circ$ & $56.34 \pm 7.89$ & $\delta_{CP}^l/^\circ$(IO) & $274_{-25}^{+22}$  & $201 \rightarrow 335$ \\ \hline
\end{tabular}
\caption{The best fit values and the $1\sigma$ uncertainties of the charged fermion mass ratios, quark mixing angles and Dirac {\it CP} violating phase of the quark sector at the GUT scale $M_{GUT} \equiv 2\times 10^{16} GeV$ with the SUSY breaking scale $M_{SUSY} = 500 GeV$ and $\tan\beta = 10$ taken from \cite{Ross}. The numerical values of the lepton mixing angles, the neutrino mass squared differences, and the leptonic {\it CP} violating phase are taken from NuFIT v6.0 with Super-Kamiokande atmospheric data for NO and IO of neutrino masses \cite{nufit}.}
\label{data}
\end{table}
The standard parametrization is employed for both the PMNS lepton mixing matrix and the CKM quark mixing matrix. In particular, the PDG parametrization of the PMNS matrix is expressed as  
\begin{equation}
U_{\text{PMNS}} = \begin{pmatrix}  
c^{l}_{12} c^{l}_{13} & s^{l}_{12} c^{l}_{13} & s^{l}_{13} e^{-i \delta^{l}_{CP}} \\  
-s^{l}_{12} c^{l}_{23} - c^{l}_{12} s^{l}_{13} s^{l}_{23} e^{i \delta^{l}_{CP}} & c^{l}_{12} c^{l}_{23} - s^{l}_{12} s^{l}_{13} s^{l}_{23} e^{i \delta^{l}_{CP}} & c^{l}_{13} s^{l}_{23} \\  
s^{l}_{12} s^{l}_{23} - c^{l}_{12} s^{l}_{13} c^{l}_{23} e^{i \delta^{l}_{CP}} & -c^{l}_{12} s^{l}_{23} - s^{l}_{12} s^{l}_{13} c^{l}_{23} e^{i \delta^{l}_{CP}} & c^{l}_{13} c^{l}_{23}  
\end{pmatrix}  
\text{diag}(1, e^{i\alpha_{21}/2}, e^{i\alpha_{31}/2}),
\label{pmns}
\end{equation}  
where $c^{l}_{ij} = \cos \theta^{l}_{ij}$, $s^{l}_{ij} = \sin \theta^{l}_{ij}$, and $\{\alpha_{21}, \alpha_{31}\}$ represent the two Majorana CP phases. In our analysis, we explore the predictions of the three $\text{PS} \times S_3$ models for the neutrino oscillation parameters, alongside several observables from non-oscillation neutrino experiments that probe the absolute neutrino mass scale. In particular, three key observables provide insights into this scale:  
\begin{itemize}
    \item The sum of active neutrino masses $\sum m_i$, constrained by cosmological observations, with the latest Planck data setting an upper limit of $\sum m_i < 0.12~\text{eV}$ \cite{Planck:2018vyg}.  
    \item The effective neutrino mass $m_\beta$, measurable in beta decay experiments by analyzing the electron energy spectrum near its endpoint. This quantity depends on neutrino masses and the elements of the first row of the PMNS mixing matrix:  
    \begin{equation}
    m_\beta = (m_1^2 \cos^2 \theta^{l}_{12} \cos^2 \theta^{l}_{13} + m_2^2 \sin^2 \theta^{l}_{12} \cos^2 \theta^{l}_{13} + m_3^2 \sin^2 \theta^{l}_{13})^{1/2}.
    \label{mb}
    \end{equation}  
    The KATRIN experiment currently provides the most robust limit on $m_\beta$, constraining the electron antineutrino mass to less than $0.45~\text{eV}$ \cite{Katrin:2024tvg}, with a future sensitivity goal of $0.2~\text{eV}$ \cite{KATRIN:2021dfa}.
    \item The effective Majorana mass $m_{\beta\beta}$, probed in $0\nu\beta\beta$ experiments. A positive signal in such experiments would also confirm the Majorana nature of neutrinos. Besides oscillation parameters, $m_{\beta\beta}$ depends on the Majorana {\it CP} phases in \ref{pmns}   
    \begin{equation}
    m_{\beta\beta} = \left| m_1 \cos^2 \theta^{l}_{12} \cos^2 \theta^{l}_{13} + m_2 \sin^2 \theta^{l}_{12} \cos^2 \theta^{l}_{13} e^{i\alpha_{21}} + m_3 \sin^2 \theta^{l}_{13} e^{i(\alpha_{31} - 2\delta^{l}_{CP})} \right|.  
    \label{mbb}
    \end{equation}  
    The KamLAND-Zen experiment currently places an upper bound on $m_{\beta\beta}$ at $m_{\beta\beta} < (28 \sim 122) \text{meV}$ \cite{KamLAND-Zen:2024eml}. Future large-scale $0\nu\beta\beta$-decay experiments aim to enhance this sensitivity, with LEGEND-1000 \cite{LEGEND:2021bnm} targeting $m_{\beta\beta} < (9 \sim 21) \text{meV}$ and nEXO \cite{nEXO:2021ujk} aiming for a sensitivity of $m_{\beta\beta} < (4.7 \sim 20.3) \text{meV}$.
\end{itemize}  
These observables, together with the unknown Majorana phases ($\alpha_{21}$ and $\alpha_{31}$) and the masses of the heavy right-handed neutrinos ($M_{i=1,2,3}$), provide testable predictions that can be utilized to assess the validity of the proposed models.
By performing a $\chi^2$ analysis, we quantify the compatibility of the benchmarks with the experimental measurements of masses and mixing angles of both leptons and quark sectors. The $\chi^2$ statistic is defined as
\begin{equation}
\label{chi2}
    \chi^2 = \sum_{i} (\frac{ P_i(\bar{x}) - \mu_{i}}{\sigma_{i}})^2,
\end{equation}
where the $\mu_{i}$ and $\sigma_{i}$ are the best fit and $\sigma$ uncertainty of the physical parameters and their values are obtained by evolving their low energy values to the GUT scale with the renormalization group equations, as shown in Table \ref{data}. The $P_i(\bar{x})$ is the prediction of the physical parameters by the model, obtained by diagonalizing the masses and mixing matrices for both leptons and quarks sectors. The $\bar{x}$ represents the model free-parameters
\begin{equation}
    \bar{x} = {a_{i}/a_{1}, b_{i}/b_{1}, c_{i}/c_{1}, r_{i}}.
\end{equation}
The total $\chi^2$ in Eq. \ref{chi2} can be divided in two terms, $\chi^2_{l}$ which refers to the $\chi^2$ involving only leptons observables and $\chi^2_{q}$ involving only quarks observables. The $\chi^2_{l}$ is constructed from lepton mass ratios and mixing angles, while the $\chi^2_{q}$ from quark mass ratios and mixing angles, as listed in Table \ref{data}. The total $\chi^2$, can be written as $\chi^2_{total} = \chi^2_{l} + \chi^2_{q}$. Note that the overall scale factors of the mass matrices do not affect the $\chi^2_{total}$ value.

\begin{figure}[h!]
    \centering
    \includegraphics[width=1\textwidth]{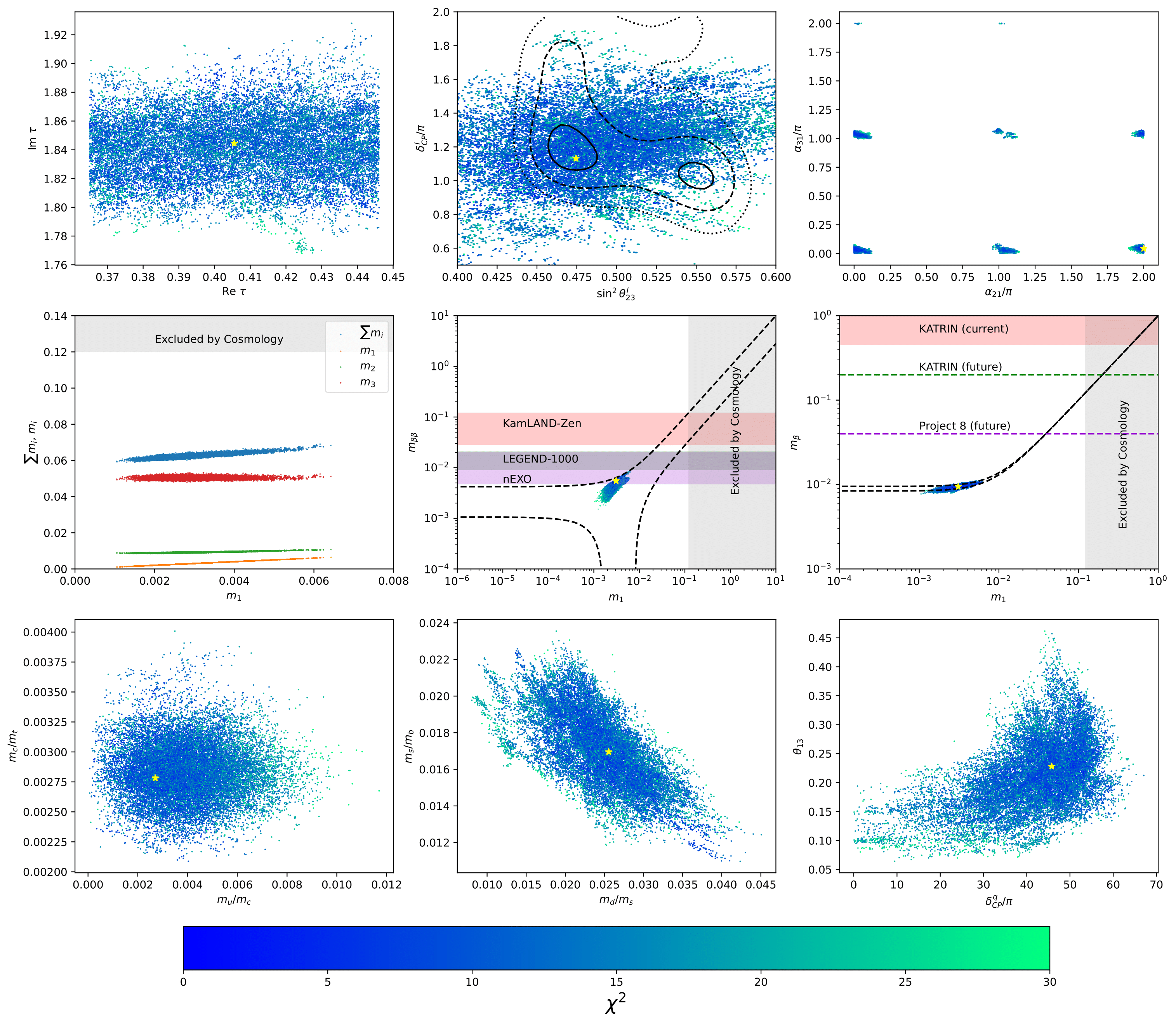}
    \caption{Allowed region of $\tau$ and predicted correlations between masses and mixing parameters of quarks and leptons in model I. The yellow star indicates the best-fitting point. Gray-shaded regions, excluded by cosmology, arise from the Planck constraint on the neutrino mass sum $\sum m_i$ \cite{Planck:2018vyg}. Shaded regions in the $m_\beta$ and $m_{\beta\beta}$ panels represent experimental limits from beta decay and $0\nu\beta\beta$ experiments, as detailed after Eqs. \ref{mb} and \ref{mbb}.}
    \label{fig1}
\end{figure}

\begin{figure}[h!]
    \centering
    \includegraphics[width=1\textwidth]{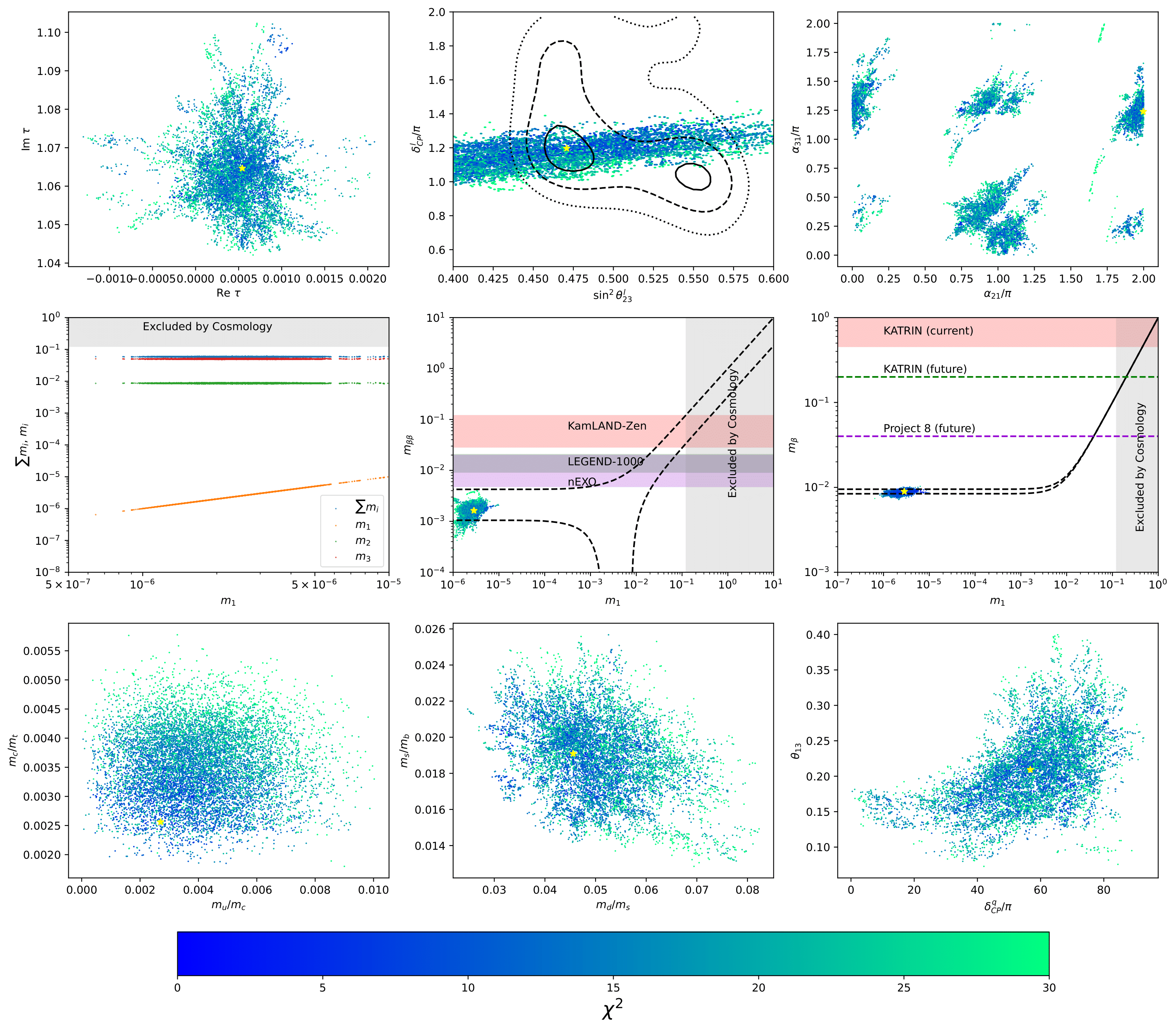}
    \caption{Same as figure 1 but for model II.}
    \label{fig2}
\end{figure}

\begin{figure}[h!]
    \centering
    \includegraphics[width=1\textwidth]{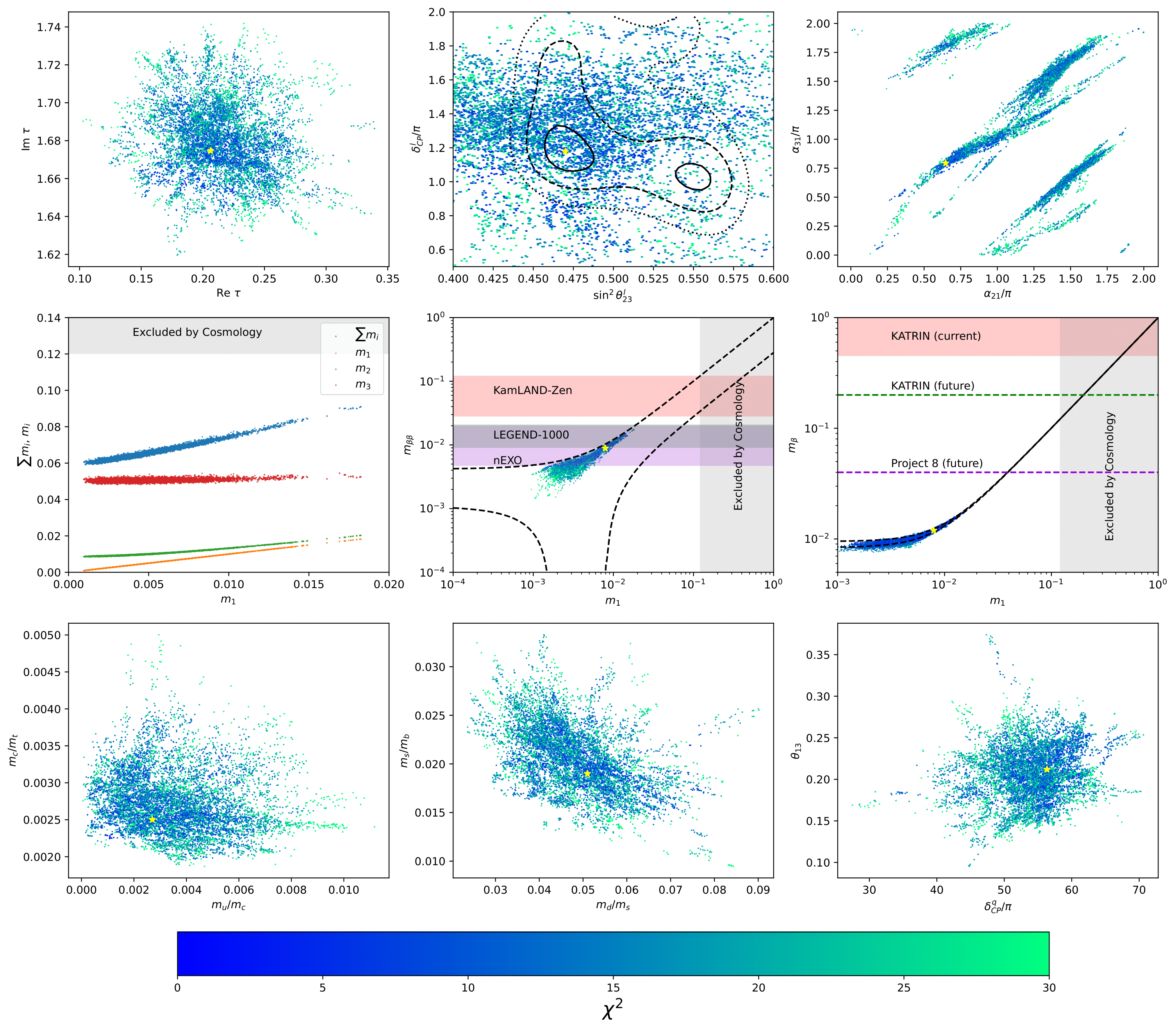}
    \caption{Same as figure 1 but for model III.}
    \label{fig3}
\end{figure}

To extract the leptons and quarks masses and mixing angles, we perform a singular value decomposition to the mass matrices, and then the $\chi^2_{total}$ is evaluated. All dimensionless parameters are treated as independent variables, with their absolute values randomly sampled within the range $[0, 10^4]$ and their phases uniformly distributed over $[0,2\pi]$. The VEV of the modulus $\tau$ is restricted to the fundamental domain, defined as ${\mathrm{Im}(\tau)>0, |\mathrm{Re}(\tau)| < 1/2, |\tau|>1}$. The global minimum is searched over the parameter space using the FlavorPy packages \cite{FlavorPy}. Due to the non-concavity of the likelihood function, the package samples a random point from the parameter space used as initial guess for the $\chi^2_{total}$ minimization. FlavorPy uses the {\it lmfit} algorithm to perform the fit. The package then perform a Markov chain Monte Carlo scan around the best-fit value. We represent our results in Table \ref{tab:comparison} and figures \ref{fig1}, \ref{fig2}, and \ref{fig3}. In particular, the best-fit values of the free parameters for the three benchmark models are provided in Table \ref{tab:comparison} for both NO and IO neutrino mass spectra. Predictions for observables, including fermion mass ratios, flavor mixing parameters, the effective Majorana neutrino mass $m_{\beta\beta}$ in $0\nu\beta\beta$, the effective electron antineutrino mass $m_\beta$, the three light neutrino masses $m_i$ and their sum $\sum m_i$, are also presented. Furthermore, the minimal values of $\chi_l^2$, $\chi_q^2$, and $\chi_{\text{total}}^2$ are reported for both mass orderings. A model is considered phenomenologically viable if it fits the 16 observables listed in Table \ref{data} within the corresponding $3\sigma$ ranges. Clearly, we see from the last row in Table \ref{tab:comparison} that the NO neutrino mass spectrum for the three benchmark models have the lowest $\chi_{total}^2$ values, and thus they are preferred over the IO\footnote{In the case of IO, Table \ref{tab:comparison} shows that two flavor observables ($\sin^2\theta^l_{13}$ and $\theta^q_{23}$) fall outside the experimental $3\sigma$ ranges for model I, and four ($m_\mu/m_\tau$, $m_d/m_s$, $m_s/m_b$ and $\delta^q_{CP}$) for model II. For model III, although only one observable ($m_s/m_b$) lies outside the $3\sigma$ allowed range, many other observables are within their $3\sigma$ ranges, making the model less predictive compared to the NO case, where all parameters lie within their $1\sigma$ allowed regions.} mass spectrum with models II and III being highly predictive. Therefore, for our numerical results we focus on the NO case. For model I, our fit results give $\chi_{\text{total}}^2 = 2.1047$, with 13 out of the 16 fitted observables lying within their $1\sigma$ experimentally allowed ranges, two within the $3\sigma$ ranges, and $m_d/m_s$ outside the $3\sigma$ range. For model II, we find $\chi_{\text{total}}^2 = 0.7581$, where all 16 fitted observables fall within their $1\sigma$ ranges. model III stands out, as our fits yield $\chi_{\text{total}}^2 = 1.8 \times 10^{-10}$, with almost all observables are at the best-fit values, as shown in the sixth column of Table \ref{tab:comparison}.
\begin{table*}
\begin{ruledtabular}
\begin{tabular}{l||c|c|c|c|c|c}
Parameters & Model I (NO) & Model I (IO) & Model II (NO) & Model II (IO) & Model III (NO) & Model III (IO) \\
\hline
$Re(\tau)$ & $0.4055$ & $-0.29218$ & $0.00054$  & $-0.2469$ & $0.2063$  & $0.4146$  \\
$Im(\tau)$ & $1.84449$ & $0.87452$ & $1.06465$  & $0.9151$ & $1.6746$  & $0.8707$  \\
$a_1 \upsilon_u~(\text{GeV})$ & $0.0028$ & $2.5119$ & $-65.9925$  & $194.2190$ & $9.8123$  & $-8.5322$
\\
$a_2/a_1$ & $2548.5970$ & $-715.4176$ & $74.5687$  & $99.9999$ & $-0.2355$ & $2.4530$  \\
$a_3/a_1$ & $35.4234$ & $-0.4506$ & $6.9255$  & $28.6055$ & $-11.8822$  & $-3.5322$  
\\
$a_4/a_1$ & --- & --- & $0.1389$ & $-3.6630$ & $46.7706$ & $28.5448$
\\
$b_1/a_1$ & $1340.01e^{1.839\pi i}$ & $112.24e^{0.280\pi i}$ & $19.38e^{0.401\pi i}$  & $0.2594e^{-0.5\pi i}$ & $59.8444e^{0.526\pi i}$ & $0.0662e^{-0.174\pi i}$
\\
$b_2/a_1$ & $1341.48e^{2.839\pi i}$ & $74.45e^{0.713\pi i}$ & $1.437e^{0.236\pi i}$ & $0.0144e^{-0.081\pi i}$ & $109.82e^{0.303\pi i}$  & $400.72e^{-0.138\pi i}$
\\
$b_3/a_1$ & $0635.17e^{4.246\pi i}$ & $133.41e^{-0.810\pi i}$ & $83.96e^{-0.227\pi i}$  &$103.52e^{1.411\pi i}$ & $979.88e^{0.598 \pi i}$  & $371.89e^{0.941\pi i}$
\\
$b_4/a_1$ & $320.63e^{2.478\pi i}$ & $2.65e^{-0.319\pi i}$ & $19.26e^{0.431\pi i}$  & $3.4e{0.046\pi i}$ & $103.26e^{0.967\pi i}$  & $1080.06e^{0.337\pi i}$
\\
$b_5/a_1$ & $542.27e^{5.945\pi i}$ & $45.10e^{-0.577\pi i}$ & $8.57e^{0.543\pi i}$ & $2.05e^{1.415\pi i}$ & $772.47e^{-0.381\pi i}$ & $931.69e^{-0.5\pi i}$
\\
$a_1 r_1 \upsilon_d~(\text{GeV})$ & $0.0033$ & $19.1259$ & $-99.8603$  & $94.3633$ & $-9.9824$  & $-9.8755$
\\
$r_2$ & $0.90e^{5.490\pi i}$ & $0.0999e^{0.344\pi i}$ & $0.941e^{-0.425\pi i}$ & $31.58e^{1.332\pi i}$ & $0.0538e^{0.891\pi i}$  & $10.64e^{0.534\pi i}$
\\
$c_2/c_1$ & $1435.74e^{2.180\pi i}$ & $27.64e^{-0.915\pi i}$ & $97.45e^{0.976\pi i}$ & $105201.9e^{-0.103\pi i}$ & $0.9047e^{-0.00188\pi i}$  & $1.0007e^{0.00041\pi i}$
\\
$c_3/c_1$ & --- & ---  & $10520.3e^{-0.603\pi i}$ & $177.03e^{1.423\pi i}$ & $104.65e^{1.204\pi i}$  & $95.46e^{0.812\pi i}$
\\
$c_4/c_1$ & --- & ---  & --- & --- & $1855.42e^{1.51\pi i}$  & $12047.13e^{0.5\pi i}$
\\
$\frac{(a_1 \upsilon_u)^2}{c_1 \upsilon_R}~(\text{meV})$ & $1.161$  & $35.759$ & $130.34$ & $100.62$ & $0.7466$  & $1.8882 \times 10^{-04}$
\\
\hline
$m_e/m_\mu$ & $0.0048$ & $0.0048$ & $0.0048$ & $0.0048$ & $0.0048$  & $0.0050$
\\
$m_\mu/m_\tau$ & $0.0551$ & $0.0616$ & $0.0587$ & $0.0513$ & $0.0590$  & $0.0550$
\\
$m_1~(\text{meV})$ & $3.086$ & $49.0925$ & $0.0028$ & $50.346$ & $7.915$  & $49.078$
\\
$m_2~(\text{meV})$ & $9.188$ & $49.849$ & $8.654$ & $51.084$ & $11.728$ & $49.835$
\\
$m_3~(\text{meV})$ & $50.227$ & $1.0087$ & $50.131$ & $11.207$ & $50.750$  & $0.0047$
\\
$\sin^2\theta^l_{12}$ & $0.306$ & $0.296$ & $0.307$ & $0.308$ & $0.308$  & $0.310$
\\
$\sin^2\theta^l_{13}$ & $0.0221$ & $0.0202$ & $0.0221$ & $0.0223$ & $0.0221$  & $0.0222$
\\
$\sin^2\theta^l_{23}$ & $0.474$ & $0.524$ & $0.470$ & $0.556$ & $0.470$  & $0.517$
\\
$\delta_{CP}^l/\pi$ & $1.131$ & $1.003$ & $1.196$ & $1.317$ & $1.178$  & $1.232$
\\
$\alpha_{21}/\pi$ & $1.999$ & $1.335$ & $1.997$ & $0.259$ & $0.647$  & $0.871$
\\
$\alpha_{31}/\pi$ & $0.043$ & $0.167$ & $1.239$ & $1.150$ & $0.793$  & $1.012$
\\
$m_\beta~(\text{meV})$ & $9.456$ & $49.86$ & $8.937$ & $51.037$ & $11.938$  & $49.787$
\\
$m_{\beta\beta}~(\text{meV})$ & $5.512$ & $42.83$ & $1.628$ & $46.743$ & $8.878$  & $44.25$
\\
\hline
$m_u/m_c$ & $0.0027$ & $0.0029$ & $0.0027$ & $0.0027$ & $0.0027$  & $0.0028$
\\
$m_c/m_t$ & $0.0027$ & $0.0025$ & $0.0025$ & $0.0028$ & $0.0025$  & $0.0020$
\\
$m_d/m_s$ & $0.0255$ & $0.0470$ & $0.0455$ & $0.02808$ & $0.0510$  & $0.0486$
\\
$m_s/m_b$ & $0.0169$ & $0.0235$ & $0.0190$ & $0.0364$ & $0.0189$  & $0.0119$
\\
$\theta^q_{12}$ & $0.229$ & $0.229$ & $0.229$ & $0.229$ & $0.229$ & $0.229$
\\
$\theta^q_{13}$ & $0.0039$ & $0.0032$ & $0.0036$ & $0.0043$ & $0.0037$ & $0.0044$
\\
$\theta^q_{23}$ & $0.0392$ & $0.0340$ & $0.0398$ & $0.0399$ & $0.0397$ & $0.0429$
\\
$\delta_{CP}^q/^\circ$ & $45.76$ & $43.93$ & $56.75$ & $93.12$ & $56.34$ & $64.55$
\\
\hline
$\chi_l^{2}$ & 0.0224 & 3.744 & 0.0377 & 1.952 & $1.8 \times 10^{-12}$ & 1.401
\\
$\chi_q^{2}$ & 2.082 & 4.041& 0.7205 & 12.742 & $1.8 \times 10^{-10}$ & 3.337
\\
$\chi_{totlal}^{2}$ & $2.1047$ & $7.7852$ & $0.7581$ & $14.694$  & $1.8\times 10^{-10}$ & $4.7385$
\end{tabular}
\end{ruledtabular}
\caption{Best-fit values of the model parameters and the corresponding predictions for fermion masses and mixing in the three benchmark PS models, invariant under the $S_3$ modular group, for both NO and IO. The numerical values of the physical observables are provided in Table \ref{data}.}
\label{tab:comparison}
\end{table*}
The allowed ranges of the modulus $\tau$ and the correlations among different observables used in our numerical fit are illustrated in figures \ref{fig1}, \ref{fig2}, and \ref{fig3}, corresponding to models I, II, and III, respectively. In each figure, the first plot shows the allowed values of the modulus $\tau$, with the best-fit values indicated by a yellow star. It is evident that $\tau$ is confined to a small, narrow region on the right side of the fundamental domain. Notably, the best-fit result for model II lies near the self-dual point $\tau=i$, where the $S$ generator of $\Gamma_2$ remains unbroken. The second plot of the first row illustrates the correlation between $\delta^l_{CP}$ and $\sin^2\theta^l_{23}$. The contours represent confidence levels (C.L.) at $1\sigma$ (solid), $2\sigma$ (dashed), and $3\sigma$ (dash-dotted). For all three models, the data predominantly lie in the lower octant of the atmospheric angle, where the best-fit values are also located: $\sin^2\theta^l_{23} = 0.474$ for model I and $\sin^2\theta^l_{23} = 0.470$ for models II and III. The corresponding best-fit values of $\delta^l_{CP}$ are $\delta^l_{CP} = 1.131\pi$ for model I, $\delta^l_{CP} = 1.196\pi$ for model II, and $\delta^l_{CP} = 1.178\pi$ for model III, all of which indicate {\it CP}-violating values. The third figure in first row displays the correlation between the two Majorana {\it CP} phases $\alpha_{21}$ and $\alpha_{31}$ where the predicted best-fit values are given by $[\alpha_{21} \sim 2\pi,~\alpha_{31}=0.043 \pi]$ for model I, $[\alpha_{21} \sim 2\pi,~\alpha_{31}=1.239 \pi]$ for model II, and $[\alpha_{21} = 0.647 \pi,~\alpha_{31}=0.793 \pi]$ for model III.
The second row presents three plots illustrating the predictions for the sum of the three neutrino masses $\sum m_i$, $m_{\beta\beta}$, and $m_\beta$ all shown as functions of the lightest neutrino mass $m_1$. The predicted best-fit values are given as follows
\begin{eqnarray}
    \text{Model I:} \quad \sum m_i &=& 62.50~\text{meV}, \quad m_{\beta\beta} = 5.512~\text{meV}, \quad m_\beta = 9.456~\text{meV}, \nonumber \\
    \text{Model II:} \quad \sum m_i &=& 58.79~\text{meV}, \quad m_{\beta\beta} = 1.628~\text{meV}, \quad m_\beta = 8.937~\text{meV}, \\
    \text{Model III:} \quad \sum m_i &=& 70.39~\text{meV}, \quad m_{\beta\beta} = 8.878~\text{meV}, \quad m_\beta = 11.938~\text{meV}. \nonumber
\end{eqnarray}
The best-fit values of $\sum m_i$ in all models are consistent with the current upper limit set by the DESI collaboration, $\sum m_i < 72~\text{meV}$ \cite{DESI:2024mwx}. Furthermore, these values lie within the sensitivity range of next-generation experiments, which are expected to probe $\sum m_i < (44-76)~\text{meV}$ using data from Euclid, CMB-S4, and LiteBIRD \cite{Euclid:2024imf}.
For models I and III, the middle panels of figures \ref{fig1} and \ref{fig3} indicate that the best-fit values of $m_{\beta\beta}$ lie below the sensitivity thresholds of both the current KamLAND-Zen experiment and the upcoming large-scale $0\nu\beta\beta$-decay experiment, LEGEND-1000. However, these values fall within the sensitivity range of the next-generation nEXO experiment, which aims to achieve $m_{\beta\beta} < (4.7 \sim 20.3)~\text{meV}$. For model II, the middle panel of figure \ref{fig2} shows that the best-fit value of $m_{\beta\beta}$ is significantly below the sensitivity of both current and future tonne-scale experiments. This is due to the lightest neutrino mass being extremely small. For instance, the best-fit value corresponds to $m_1 = 0.002845~\text{meV}$. The effective electron antineutrino mass $m_\beta$ is illustrated in the last panel of the second row in figures \ref{fig1}, \ref{fig2}, and \ref{fig3}, corresponding to models I, II, and III, respectively. Across all viable models, the best-fit values of $m_\beta$ remain below the projected sensitivity of $40~\text{meV}$ anticipated by the future Project 8 experiment.
The last three panels of each figure depict the correlations among quark mass ratios as well as between the quark Dirac {\it CP} phase and the third quark mixing angle. In model I, the optimal value for $m_d/m_s$ lies outside the $3\sigma$ range, while all other parameters fall within the $1\sigma$ range. For models II and III, the best-fit values for all parameters remain well within the $1\sigma$ range.

Finally, we note that our results remain in a good agreement with a SUSY-breaking scale $M_{SUSY}$ of $1~TeV$. The best-fit values and $1\sigma$ uncertainties for the fermion mass ratios, the quark mixing angles and the{\it CP} violating phase observables at the GUT scale $2 \times 10^{16}~GeV$, assuming $\tan\beta = 10$ and $M_{SUSY} = 1~TeV$, are taken from Ref. \cite{Antusch:2013jca}--see also Ref. \cite{Bjorkeroth:2015ora} for a summary of results at $\tan\beta = 5, 10$-- and presented as follows:
\begin{eqnarray}
    \frac{m_e}{m_\mu} &=& 0.00472 \pm 0.00008, \quad  \frac{m_\mu}{m_\tau} = 0.05860 \pm 0.00092, \quad \frac{m_u}{m_c} = 0.00204 \pm 0.00127, \quad \frac{m_c}{m_t} = 0.00271 \pm 0.00025, \nonumber \\
    \frac{m_d}{m_s} &=& 0.00472 \pm 0.00008, \quad  \frac{m_s}{m_b} = 0.05046 \pm 0.01242, \quad \frac{m_u}{m_c} = 0.01368 \pm 0.00152,   \\
    \theta_{12}^q &=& 0.22736 \pm 0.00142, \quad \theta_{13}^q = 0.00314 \pm 0.00049, \quad \theta_{23}^q = 0.03584 \pm 0.00670, \quad \delta_{CP}^q/^\circ = 69.21 \pm 6.19. \nonumber
\end{eqnarray}
Using this numerical values, along with the neutrino oscillation data in Table \ref{data}, we find that our fit in the NO case yields $\chi_{total}^2 = 5.6179$ for model I, $\chi_{total}^2 = 2.3143$ for model II, and $\chi_{total}^2 = 2.2928$ for model III. In model I, only two out of the 16 fitted observables fall outside their $3\sigma$ range: the lepton mass ratio $m_\mu / m_\tau = 0.05517$ and the Dirac {\it CP} phase from the quark sector, $\delta_{CP}^q/^\circ = 45.76$. In model II, the only parameter outside the $3\sigma$ range is the mass ratio $m_s/m_b$, with a best-fit value of $0.019086$, whereas $\delta_{CP}^q/^\circ = 56.75^\circ$ remains within the $3\sigma$ range. Model III exhibits a similar pattern, with $m_s/m_b = 0.01899$ lying slightly outside the $3\sigma$ range, while $\delta_{CP}^q/^\circ = 56.345^\circ$ remains consistent with the data.  
Models achieving $\chi_{total}^2 < 10$ can be considered in good agreement with the data, given the large number of observables that are fitted. However, further refinements--such as introducing corrections to the superpotential that modify the fermion mass matrices--could improve the fit even further, potentially achieving a perfect fit such as the one obtained for model III in the NO case, as reported in Table \ref{data}.
\section{Summary and discussion}
\label{concl}
Implementing non-Abelian discrete flavor symmetries into GUTs provides a compelling approach to addressing the flavor structure of quarks and leptons. On the other hand, modular invariance has emerged as a promising alternative to conventional flavor symmetries, sidestepping the challenges associated with flavon vacuum alignment and reducing the free parameters by describing Yukawa couplings as modular forms that depend on a single complex modulus $\tau$. In this work, we have explored the fermion flavor structure within the Pati-Salam GUT framework, incorporating the $S_3$ modular group for the first time. We have examined three benchmark models, which differ based on the transformation properties of the matter fields $F_i \sim (4,2,1)$ and $F_i^c \sim (\Bar{4},1,2)$ under the modular symmetry $S_3$ as well as their corresponding modular weights. In the scalar sector, we have adopted a minimal field content consisting of the multiplets $\Phi = (1, 2, 2)$, $\Sigma = (15, 2, 2)$, and $\Delta_R = (10, 1, 3)$. The VEV of $\Delta_R$ breaks the PS gauge group down to the SM gauge group and generates a Majorana mass for the RH neutrinos. Meanwhile, the VEVs of $\Phi$ and $\Sigma$ further break the SM gauge group to $SU(3)_C \otimes U(1)_Q$ and provide masses for the down quarks, up quarks, charged leptons, and Dirac neutrinos. In our benchmark models, all three multiplets are assumed to transform trivially under the modular $S_3$ group and are assigned identical modular weights.

We have performed a comprehensive numerical analysis for each model, investigating the allowed values of the modulus $\tau$, correlations among observables, and predictions for lepton and quark parameters. Our findings reveal that all models favor the NO mass spectrum, with the best-fit values of $\tau$ confined to narrow regions within the fundamental domain. The models predict {\it CP}-violating Dirac and Majorana phases, with the atmospheric mixing angle $\sin^2\theta^l_{23}$ favoring the lower octant. 
The neutrino mass parameters $\left(\sum m_i, m_{\beta\beta}, m_\beta\right)$ align with current experimental constraints and provide insights into upcoming detection capabilities. The sum of neutrino masses, $\sum m_i$, satisfies the DESI limit of $\sum m_i < 72~\text{meV}$ and falls within the sensitivity range of future experiments. Regarding $m_{\beta\beta}$, model II predicts values below the detection thresholds of upcoming experiments, whereas models I and III offer predictions accessible to the nEXO experiment. Similarly, for $m_\beta$, all models yield predictions below the projected sensitivity of the Project 8 experiment.  In the quark sector, the predicted mass ratios and mixing parameters generally agree with experimental data. However, a minor tension arises in model I, where the ratio $m_d/m_s$ slightly exceeds the $3\sigma$ experimental range.

One of the most distinctive features of the PS model is its natural inclusion of RH neutrinos, which play a crucial role in explaining the smallness of neutrino masses via the seesaw mechanism. In this framework, the seesaw scale is directly linked to the symmetry-breaking scale, determined by the VEV of $\Delta_R$, denoted as $\langle \Delta_R \rangle = \upsilon_R$. From our numerical analysis, we estimate this scale by considering the overall mass scale of the neutrino mass matrix, derived using the Type I seesaw formula, expressed as $(a_1 \upsilon_u)^2 / c_1 \upsilon_R$. By using the numerical values of this overall mass scale and $a_1 \upsilon_u$ provided in Table~\ref{tab:comparison}, we calculate the approximate PS symmetry breaking scales for each model where we find
\begin{eqnarray}
    \text{Model I (NO):} \quad c_1 \upsilon_R = 7.89 \times 10^{3}~\text{GeV}, \nonumber \\
    \text{Model II (NO):} \quad c_1 \upsilon_R = 3.34 \times 10^{13}~\text{GeV}, \\
    \text{Model III (NO):} \quad c_1 \upsilon_R = 1.29 \times 10^{14}~\text{GeV}. \nonumber
\end{eqnarray}
Interestingly, assuming $c_1 \sim \mathcal{O}(1)$, model I corresponds to a low-scale PS scenario, while models II and III align with high-scale PS scenario. With these variations in PS models and the numerical parameters summarized in Table~\ref{tab:comparison}, we diagonalize the Majorana mass matrix for each model. The resulting eigenvalues correspond to the masses of the three RH neutrinos
\begin{eqnarray}
    \text{Model I (NO):} \quad M_1 &\sim& 1.0 \times 10^3~\text{GeV}, \quad M_2 \sim 1.4 \times 10^{6}~\text{GeV}, \quad M_3 \sim 1.4 \times 10^{6}~\text{GeV}, \nonumber \\
    \text{Model II (NO):} \quad M_1 &\sim& 3.3 \times 10^{13}~\text{GeV}, \quad M_2 \sim 3.25 \times 10^{15}~\text{GeV}, \quad M_3 \sim 7.15 \times 10^{15}~\text{GeV}, \\
    \text{Model III (NO):} \quad M_1 &\sim& 1.9 \times 10^{11}~\text{GeV}, \quad M_2 \sim 8.7 \times 10^{12}~\text{GeV}, \quad M_3 \sim 2.4 \times 10^{17}~\text{GeV}. \nonumber
\end{eqnarray}
These masses provide a pathway to addressing a fundamental challenge in particle physics: the baryon asymmetry of the universe. In particular, the inclusion of RH neutrinos enables the explanation of this asymmetry through the leptogenesis mechanism \cite{Fukugita:1986hr}. Moreover, when RH neutrinos are present in a model, the running effects should be considered from the cutoff scale down to the mass of the lightest heavy neutrino $M_1$ as described in Ref. \cite{Criado:2018thu}. Investigating the viability of
leptogenesis and the effects of RGE in each scenario in the different models requires a thorough and in-depth analysis, which extends beyond the scope of this paper.  We therefore defer this investigation to future work.

\section*{Acknowledgment}
This work is supported by the United Arab Emirates University (UAEU) under UPAR Grant No. 12S162 and Start-Up Grant No 12S157.
\appendix
\section{$S_3$ tensor product rules and higher weight modular forms}
\label{appe}
The group $S_3$ consists of all permutations of a three-element set and corresponds to the symmetries of an equilateral triangle, encompassing three rotations and three reflections. It is the smallest non-Abelian discrete group and consists of $3!=6$ permutations, thereby containing six elements. This group can be generated by two elements $S$ and $T$. Following the convention established in \cite{Ishi}, we will represent these generators using a real basis, given by the following symmetric matrices
\begin{eqnarray}
    S = -\frac{1}{2} \begin{pmatrix}
        1 & \sqrt{3} \\
        \sqrt{3} & -1
    \end{pmatrix} , \quad T = \begin{pmatrix}
       1 & 0 \\
       0 & -1
    \end{pmatrix},
\end{eqnarray}
satisfying $S^2 = T^2 = (ST)^3 = 1$. Using the standard relation that connects the order of $S_3$ with the dimensions of its irreducible representations, $6 = 1^2 + 1^2 + 2^2$, we deduce that $S_3$ has three irreducible representations: two singlets, $R_1 \equiv \bm{1}$ (trivial) and $R_{1^{\prime}} \equiv \bm{1^\prime}$ (pseudo-singlet), and one doublet, $R_2 \equiv \bm{2}$. \newline
Let us now provide the tensor decomposition rules of the irreducible representations of the $S_3$ group. Let us denote the pseudo-singlets by $z_i$ and take two $S_3$ doublets with components $x_i$ and $y_i$ for $i=1,2$. The tensor product rules are given as follows
\begin{eqnarray}
    R_{1^{\prime}} &\otimes& R_{1^{\prime}} = R_1 \sim z_1 z_2 , \quad  R_{1^{\prime}} \otimes R_2 = R_2 \sim \begin{pmatrix}
        -z_1 x_2 \\ ~~z_1 x_1
    \end{pmatrix} \nonumber \\
    R_2 &\otimes& R_2 = R_1 \oplus R_{1^{\prime}} \oplus R_2 
    \left\{ 
    \begin{array}{c}
     R_1 \sim x_1 y_1 + x_2 y_2 \\ 
     R_{1^{\prime}} \sim x_1 y_2 - x_2 y_1 \\ 
     R_2 \sim \begin{pmatrix}
         x_2 y_2 - x_1 y_1 \\
         x_1 y_2 + x_2 y_1
     \end{pmatrix}
    \end{array}
\right. 
\label{A2}
\end{eqnarray}
Using these tensor products, all higher-weight modular forms of level 2 can be constructed. Thus, following the discussion of section \ref{sec3}, the modular forms of weight $k=4$ are obtained by taking the tensor product of two basic weight-2 modular forms doublets, $Y_2^{(2)}$. Specifically, the tensor product of two weight-2 forms is given by
\begin{eqnarray}
Y_{\bm{2}}^{(2)}(\tau) \otimes Y^{(2)}_{\bm{2}}(\tau) = \left( Y_{\bm{1}}^2(\tau) + Y_{\bm{2}}^2(\tau) \right)_{\bm{1}} \oplus \left( Y_{\bm{2}}^2(\tau) - Y_{\bm{1}}^2(\tau), \, 2Y_{\bm{1}}(\tau)Y_{\bm{2}}(\tau) \right)_{\bm{2}}.
\end{eqnarray}
This decomposition yields three modular forms of weight 4 and level 2: a singlet $\bm{1}$, expressed as $Y^{(4)}_{\bm{1}}(\tau) = Y_{\bm{1}}^2(\tau) + Y_{\bm{2}}^2(\tau)$, and the other two forming an $S_3$ doublet $\bm{2}$, given by $Y^{(4)}_2(\tau) = \left( Y_{\bm{2}}^2(\tau) - Y_{\bm{1}}^2(\tau), \, 2Y_{\bm{1}}(\tau)Y_{\bm{2}}(\tau) \right)^T$. Note that the pseudo-singlet modular form of weight 4 vanishes due to the decomposition of the tensor product of two $S_3$ doublets, as shown in Eq. \ref{A2}. For modular forms of weight $k=6$ and level 2, there are four independent modular forms, which can be constructed by using the tensor product of a weight-2 modular form $Y_{\bm{2}}^{(3)}$ and a weight-2 modular form $Y_{\bm{2}}^{(4)}$. The decomposition of this tensor product gives rise to the following singlet, pseudo-singlet, and doublet weight 6 modular forms
\begin{eqnarray}
Y_{\bm{1}}^{(6)}(\tau) &=& 3Y_{\bm{1}}(\tau)Y_{\bm{2}}^2(\tau) - Y_{\bm{1}}^3(\tau), \nonumber \\
Y_{\bm{1^\prime}}^{(6)}(\tau) &=& Y_{\bm{2}}^3(\tau) - 3Y_{\bm{2}}(\tau)Y_{\bm{1}}^2(\tau), \\
Y^{(6)}_{\bm{2}}(\tau) &=& \left( Y_{\bm{1}}(\tau)(Y_{\bm{1}}^2(\tau) + Y_{\bm{2}}^2(\tau)), \, Y_{\bm{2}}(\tau)(Y_{\bm{1}}^2(\tau) + Y_{\bm{2}}^2(\tau)) \right). \nonumber
\end{eqnarray}
These modular forms serve as the building blocks for constructing Yukawa couplings and other interaction terms in models based on $\Gamma_2 \equiv S_3$. The higher-weight ($k>6$) modular forms expand the possibilities for more complex model-building scenarios, providing multiple options for coupling different representations in modular-invariant theories.

\bibliographystyle{JHEP}
\bibliography{bibliography.bib}

\end{document}